\documentclass[11pt,a4paper]{article}
\usepackage{amsmath}
\usepackage{amsfonts}
\usepackage{multirow}
\usepackage{graphicx}
\usepackage{float}
\usepackage[colorlinks=true,linkcolor=blue]{hyperref}

\newcommand{\be}{\begin{equation}}
\newcommand{\ee}{\end{equation}}
\newcommand{\bea}{\begin{eqnarray}}
\newcommand{\eea}{\end{eqnarray}}
\newcommand{\half}{\frac{1}{2}}
\newcommand{\pr}{\partial}
\newcommand{\M}{{\cal M}}
\newcommand{\eps}{\varepsilon}
\newcommand{\R}{\hbox{\upright\rlap{I}\kern 1.7pt R}}

\title{\textbf{Collective Coordinate Models for 2-Vortex Shape Mode Dynamics}}

\author{A. Alonso Izquierdo$^{(a)}$, N. S. Manton$^{(b)}$,
J. Mateos Guilarte$^{(c)}$ and A. Wereszczynski$^{(a,d,e)}$
\\ 
{\normalsize {\it $^{(a)}$ Departamento de Matematica
Aplicada}, {\it Universidad de Salamanca, SPAIN}}\\
{\normalsize {\it $^{(b)}$ Department of Applied Mathematics and Theoretical Physics}, {\it University of Cambridge, UK}} \\
{\normalsize {\it $^{(c)}$ Departamento de Fisica Fundamental}, {\it
    Universidad de Salamanca, SPAIN}}
    \\
{\normalsize {\it $^{(d)}$ Institute of Theoretical Physics}, {\it
   Jagiellonian University, POLAND}}
    \\
{\normalsize {\it $^{(e)}$ International Institute for Sustainability with Knotted Chiral Meta Matter (WPI-SKCM2)}}, \\
{\normalsize {\it
    Hiroshima University, JAPAN}}}

\date{}

\begin{document}

\maketitle

\begin{abstract}
Models are developed for the motion of charge-2 Abelian Higgs vortices
through the 2-vortex moduli space $\M$, with the vortices excited by their
shape mode oscillations. The models simplify to the well-known
geodesic flow on $\M$, modified by a potential, when the mode oscillations
are fast relative to the moduli space motion and their amplitudes are
small. When the lowest-frequency mode is excited with a large
amplitude, the geodesic flow is not a correct description. Instead, a
chaotic, or even fractal, multi-bounce structure in vortex-vortex
collisions is predicted.
\end{abstract}

\section{Introduction}

It has been known for some time that a unit-charge, critically coupled
Abelian Higgs vortex (a BPS vortex) has a unique shape mode -- a
discrete, normalizable,
radial oscillation mode whose frequency is below that of the continuum
of radiation modes \cite{Arodz1}. More recently, the shape modes of
higher-charge, circularly symmetric (i.e. coincident) vortices have been
determined. There are a finite number of these modes, the number
increasing somewhat irregularly with the charge
\cite{Hindmarsh,AlGarGuil,AlGarGuil2}. More recently still, the present
authors studied how these shape modes and their frequencies vary over the
2-vortex moduli space \cite{AGMM}. We confirmed that the
circularly symmetric 2-vortex has three shape modes, of which the
upper two are degenerate. We also found that the degeneracy is broken
as the vortices separate, and at a modest separation the
highest mode disappears into the continuum spectrum. The
remaining two modes, as the vortices separate further, are the in-phase and
$180^\circ$ out-of-phase combinations of the radial shape modes of the two
individual vortices. The in-phase combination has the lower
frequency, but the frequencies merge at asymptotically large separation.

Over the 2-vortex moduli space, then, there is a non-degenerate
lowest-frequency shape mode, and two higher-frequency
modes, whose frequencies degenerate at vortex
coincidence. Here, we first construct a model for the oscillatory
dynamics of the lowest mode coupled to 2-vortex motion through
the moduli space, and then a model for the dynamics of the two
higher-frequency modes. We do not attempt to model a simultaneous
excitation of all three modes.

We will need to use the curved metric on the 2-vortex moduli
space. Samols investigated this metric for vortices moving in a plane,
and discovered a formula for it just involving field data close to the
vortex centres. This could be exploited to calculate the metric
numerically \cite{Samols}. The
centre of mass decouples and has a flat metric, so the nontrivial factor is
the moduli space $\M$ for the relative motion of the two
vortices, with the centre of mass fixed. $\M$ is a
smooth, two-dimensional manifold with $O(2)$ rotational symmetry -- a
surface of revolution -- and its curvature is such that it can be
embedded in $\mathbb{R}^{3}$ as a rounded cone whose apex corresponds
to coincident vortices. The mode frequencies
vary over $\M$, respecting this rotational symmetry.

The model for the lowest shape mode is simpler than the model for the
upper modes. Its ingredients are the metric and the lowest mode's
frequency over $\M$.
The resulting three-variable dynamics is not integrable, but provided
the mode amplitude is small, and the motion through $\M$ is
slow, we can perform an adiabatic analysis, because the
mode oscillation is fast compared with the motion through $\M$.
The reduced, two-variable dynamics is then integrable, using
the conserved angular momentum and conserved energy. We find that the
geodesic dynamics through moduli space that occurs in the absence of the
mode excitation is modified by an induced potential energy
function proportional to the mode's frequency. Such a result is
standard in the context of adiabatic dynamics.

The second model, for the excited upper pair of modes, is more
sophisticated because of the conical structure of the frequencies around
the point of vortex coincidence (the apex of $\M$), but it is only applicable
when the two vortices are close together, as the highest mode (the
third) disappears into the continuum spectrum when the vortices reach
a modest separation. We can therefore approximate the
moduli space as being flat, or of constant curvature, because its
relevant inner region is a small neighbourhood of the
apex. The generic dynamics is still quite complicated, and is
best studied numerically.

This model simplifies if the relative vortex motion is
restricted to be head-on, and it naturally joins up to a model
for head-on motion in the outer region of moduli space, where the
third mode is absent. We can therefore study a head-on collision where
the vortices approach from infinity with the second mode excited. In
the adiabatic approximation, this is similar to a head-on collision
with the lowest mode excited, but the induced potential is repulsive rather
than attractive.

Both the metric and mode frequencies over the moduli space $\M$ are
only known numerically, but it is helpful to work with analytical
formulae. We start in section 2, therefore, by deriving a good
analytical approximation to the metric (and a simplified variant of
this), and also present simple formulae for the frequencies of the
three modes that fit the numerical results.

Section 3 discusses the model for the dynamics with the lowest mode
excited, and presents numerical results for a head-on collision.
The main change, in comparison with the geodesic flow
describing the evolution of unexcited BPS vortices, is the appearance
of an attractive mode-induced force. This has a great effect on
vortex-vortex collisions. When the amplitude of oscillation is fairly large,
the dynamics is complicated and exhibits a chaotic structure of
multi-bounce windows reminiscent of what occurs in kink-antikink
dynamics in one space dimension \cite{sug, CSW, MORW}. In this
regime the geodesic dynamics fails completely. We also perform
the small-amplitude adiabatic analysis and derive a reduced dynamics
that is integrable. Here a head-on vortex collision always leads
to scattering through $90^\circ$ (i.e. one bounce) and there are no
multi-bounces.

Section 4 discusses the dynamics with the higher-frequency modes
excited. Here, the adiabatic dynamics for small-amplitude
oscillations is more interesting, because of the degeneracy
of the second and third modes when the vortices coincide.
Large-amplitude oscillations lead to complicated dynamics that we do not
study in any detail. However we do present numerical results for
head-on collisions where only the second mode is initially
excited, and the excitation transfers to the third mode if the vortices
pass through the circularly symmetric configuration at the apex of
the moduli space, and scatter through $90^\circ$. In this case, the
excitation of the mode gives rise to a force which changes sign at the
apex. An initially repulsive interaction as the vortices approach
changes into an attractive one as they separate at $90^\circ$. As a
consequence, the vortices can stop and return -- there is then
2-bounce (i.e. backwards) scattering. No further bounces are observed
to occur.

In an appendix, we present some details of how our finite-dimensional
models for the modes, coupled to the moduli space dynamics, emerge
from the underlying field theory.

\section{Approximate metric and mode frequencies
on the 2-vortex moduli space}

The lowest-order dynamics of two unit-charge BPS vortices is captured
by geodesic motion on the moduli space $\M$ equipped with
a curved metric originally found by Samols \cite{Samols}. Here, the
kinetic degrees of freedom of the vortices are excited but their
internal shape modes are not.

In the physical 2-plane we use Cartesian coordinates $(X_1,X_2)$. The
Higgs field $\phi$ of a centred 2-vortex has zeros at an unconstrained pair
of locations $(X_1,X_2) = \pm (x_1,x_2)$, so we can denote a point in
$\M$ (up to a sign) by Cartesian/polar coordinates $(x_1,x_2)
= (\rho \cos\theta, \rho \sin\theta)$, and combine these into the
complex coordinate $w = x_1 + ix_2 = \rho e^{i\theta}$. We refer to
the real and imaginary axes in the $w$-plane (and also the $X_1$- and
$X_2$-axis in the physical plane) as horizontal and vertical,
respectively. $w$ is a natural complex coordinate
on $\M$, with its magnitude and argument having the ranges $\rho \geq 0$
and $\theta \in [-\half\pi,\half\pi]$. The vortex centres (the
Higgs field zeros) are precisely at $w$ and $-w$ in the physical 2-plane.
Because vortices are indistinguishable, a shift of $\theta$ by $\pi$ maps
a 2-vortex configuration into itself, which explains the limited range of
$\theta$. A simple geodesic on $\M$ is where the vortices approach
head-on along the horizontal axis, instantaneously coalesce at the
origin, and then separate along the vertical axis. Here, $\theta$
jumps by $\half \pi$.

The exact metric on $\M$ has the general, circularly symmetric form
\be
ds^2=f^2(\rho) \left( d\rho^2 + \rho^2 d\theta^2 \right).
\label{metric}
\ee
For small and large $\rho$, the conformal factor is
\be
f^2(\rho) = \left\{
\begin{array}{cc}
2\pi \gamma \rho^2 & \rho \to 0 \,,\\
2\pi & \rho \to \infty \,.
\end{array}
\right.
\ee
Here, we have absorbed a factor of $2\pi$ into Samols' original
metric; $\gamma$ is approximately 0.433. For large $\rho$, the metric
is asymptotically flat, with an exponentially small correction \cite{MSp}
that we neglect. Because of the range of $\theta$, the moduli space $\M$
is not asymptotically a plane, but an intrinsically flat cone, whose
half-opening angle is $30^\circ$ (in the embedding in $\mathbb{R}^{3}$).
$\M$ is globally a rounded cone -- a flat cone whose apex is smoothly
rounded off. The factor $\rho^2$ ensures that the metric is smooth at $\rho=0$.
This is verified by changing to a coordinate $z$ proportional to $w^2$. $z$,
whose argument has range $2\pi$, is a better global coordinate than
$w$ on $\M$, as it ignores the sign of $w$, thereby taking into
account the identity of the two vortices. If we write $z = x+iy
= re^{i\varphi}$, then $x=r\cos\varphi$ and $y=r\sin\varphi$ are useful
Cartesian coordinates on $\M$, even though $\M$ is curved. Then,
\be
ds^2=\Omega(x,y)(dx^2+dy^2) = \Omega(r)(dr^2 + r^2 d\varphi^2) \,,
\ee
where the conformal factor $\Omega$ is a function only of
$r=\sqrt{x^2+y^2}$.

A head-on collision, described earlier using the coordinate $w$,
becomes simply a smooth motion along the $x$-axis from $+\infty$ to $-\infty$.
However, because the geodesics on the asymptotically flat cone are simply
straight lines in terms of $w$, we will often work with $w$
as the coordinate on $\M$, but at other times with $z$.

For our purposes, it is convenient to have good approximations to the metric
on $\M$, enabling us to determine geodesics
analytically, and hence vortex scattering in the absence of shape mode
oscillations. To test these approximations, we compare the dependence
of the scattering angle on impact parameter with the dependence
obtained for the exact metric, presented graphically by Samols
\cite{Samols}. We consider two approximate metrics and will use
these later when discussing small-amplitude oscillations of the vortex
shape modes, and their effects on vortex dynamics and scattering.
These approximations to the Samols geometry are novel and could be more
broadly useful, e.g. for studying vortex quantum states, or
approximating the moduli space metric for more than two vortices.

A striking result of Samols is that the rounded cone
has an area deficit of $2\pi^2$ relative to the completed flat cone,
extrapolated to its pointed apex \cite{Samols}. We will reproduce this area
deficit exactly with our approximate metrics.

\subsection{Spherical cap approximation}

For our first approximation, we attach a spherical cap to a truncated
flat cone of opening angle $30^\circ$, maintaining a continuous tangent.
The join needs to be at $\rho = \rho_0 = \sqrt{6}$. The total metric is
\be
ds^2_{(1)}=f^2_{(1)}(\rho) \left( d\rho^2 + \rho^2 d\theta^2 \right) \,,
\label{app1met}
\ee
where
\be
f^2_{(1)}(\rho) = \left\{
\begin{array}{cc}
\frac{6912 \pi \rho^2}{(108 +\rho^4)^2} & \rho \leq \sqrt{6} \,, \\
& \\
2\pi & \rho \geq \sqrt{6} \,.
\end{array}
\right.
\label{app1metric}
\ee
$f^2_{(1)}(\rho)$ is continuous, and both functions in
(\ref{app1metric}) have zero derivative at $\rho = \sqrt{6}$.
To verify that for $\rho \leq \sqrt{6}$ it is a sphere metric, we
introduce\footnote{$w = \rho e^{i\theta}$ is used consistently throughout
this paper; $z = kw^2$ for some constant positive multiple $k$, but
$k$ varies between (sub)sections.}
\be
z=\frac{1}{\sqrt{108}}w^2=\frac{1}{\sqrt{108}} \rho^2 e^{i2\theta} \,.
\ee
In terms of $z$ we find that, for $\rho \leq \sqrt{6}$,
\be
ds^2_{(1)} = \frac{6912 \pi \rho^2}{(108 +\rho^4)^2}  \left( d\rho^2 +
\rho^2 d\theta^2 \right) = \frac{16\pi \, dz d{\bar z}}{(1 + z{\bar
  z})^2} \,.
\ee
The last expression represents the metric on a sphere with
squared radius $4\pi$, with $z$ the stereographic coordinate.

Note that the spherical cap is restricted to $\rho \leq \sqrt{6}$
which implies that $|z| \leq \frac{1}{\sqrt{3}}$. Hence, on the cap,
the maximal polar angle is $\mu = 60^\circ$. (Use $|z| = \tan \half \mu$ to
verify this.) The boundary circles of the cap and the truncated cone
have equal lengths $\sqrt{12 \pi^3}$. Also $|z|$ has the correct
range for the cap to join the flat cone with a continuous tangent.

Finally, the area deficit of the approximate metric (\ref{app1met})
is correct. To see this we compare the area of the missing part of
the flat cone,
\be
A_{\rm{cone}} = \pi \int_0^{\sqrt{6}} 2\pi  \rho \, d\rho = 6\pi^2 \,,
\ee
where the prefactor is the range of $\theta$, with the area of the
spherical cap
\be
A_{\rm{cap}} = \pi \int_0^{\sqrt{6}}
\frac{6912\pi \rho^3}{(108 + \rho^4)^2} \, d\rho = 4\pi^2 \,.
\ee
$A_{\rm{cap}}$ is one quarter of the area of a complete sphere of squared
radius $4\pi$. The difference of the areas is
\be
A_{\rm{cone}} - A_{\rm{cap}} = 2\pi^2 \,,
\ee
as required.

In this spherical cap approximation, a geodesic on $\M$ is formed
from a straight line on the flat cone, joined to a segment of a
great circle on the spherical cap, joined to another straight line on the
cone. It will be convenient to consider the geodesics in the right-hand
half $w$-plane that are reflection-symmetric with respect to the real
axis. Any geodesic can be rotated to such a position.
Some geodesics on the cone are sufficiently far from the vertex that they
do not intersect the spherical cap. These geodesics are complete
straight lines in the $w$-plane, parallel to the imaginary axis,
describing 2-vortex motion without scattering.

For the approximate metric (\ref{app1met}), with the conformal factor
$f^2_{(1)}(\rho)$, a dynamical geodesic trajectory
$w(t) = \rho(t) e^{i\theta(t)}$ arises as the solution of the equation
of motion for a particle with Lagrangian
\be
L = \half f^2_{(1)}(\rho)(\dot\rho^2 + \rho^2 \dot\theta^2) \,.
\ee
There are two constants of motion, the energy $E$ and angular momentum
$J$. $E$ is the same expression as $L$, and
$J = f^2_{(1)}(\rho) \rho^2 \dot\theta$.

Asymptotically, $f^2_{(1)}(\rho) = 2\pi$, as $\pi$ is the mass of
a 1-vortex. The angular momentum of the incoming motion is $J = 2\pi
v_{\rm in} a$, where $v_{\rm in}$ is the speed of each vortex and
$a$ the impact parameter (half the orthogonal separation of the
incoming, parallel paths of the two vortices in the physical plane).
The initial energy is $E = \pi v_{\rm in}^2$,
so $J^2/E = 4\pi a^2$. Any geodesic has a point of closest
approach of the two vortices, where $\rho$ takes its minimum value
$\tilde\rho$. Here $\dot\rho = 0$, so
$E = \half f^2_{(1)}(\tilde\rho) \tilde\rho^2 \dot\theta^2$
and $J = f^2_{(1)}(\tilde\rho) \tilde\rho^2 \dot\theta$. $\dot\theta$
cancels in $J^2/E$, and conservation of $J^2/E$ implies that
\be
f^2_{(1)}({\tilde\rho}) {\tilde\rho}^2 = 2\pi a^2 \,.
\label{closest}
\ee
This relation between the closest approach and the impact parameter will
be useful shortly. For a geodesic that does not intersect the
spherical cap, $\tilde\rho$ is simply $a$.

Geodesics on the spherical cap are segments of great circles on the
complete sphere. In terms of the stereographic coordinate $z$ on the
cap, these great circles are a family of circles in the $z$-plane,
having the algebraic equation
\be
z\bar{z} + \beta(z+\bar{z}) - 1 = 0 \,,
\label{circle}
\ee
where the parameter $\beta$ is real and positive. This family is
algebraically the linear join of the equatorial great circle
$z\bar{z} - 1 = 0$ and the line $z + \bar{z} = 0$, which describes a
great circle through the pole of the cap (the rounded cone's apex).
They are all great circles because they pass through the antipodal
points $i$ and $-i$ on the sphere. 
As $|z| \le \frac{1}{\sqrt{3}}$ on the spherical cap, the relevant
range of $\beta$ is $\frac{1}{\sqrt{3}} \le \beta \le \infty$. For
$\beta = \frac{1}{\sqrt{3}}$ the geodesic just touches the
cap at $z = \frac{1}{\sqrt{3}}$, and for $\beta = \infty$ the geodesic
passes through the pole, and represents a head-on collision of vortices.

For the conformal factor on the spherical cap (\ref{app1metric}), the
relation giving the closest approach is
\be
\frac{6912\pi{\tilde\rho}^4}{(108 + {\tilde\rho}^4)^2} = 2\pi a^2 \,,
\ee
so
\be
\tilde\rho^2 = \frac{12\sqrt{6}}{a} \left( 1 - \sqrt{1 -
    \frac{a^2}{8}} \right) \,.
\ee
From this, we can determine the relation between $\beta$ and the impact
parameter $a$. The closest approach of the circle (\ref{circle})
to the origin is where the circle crosses the real axis. This is where
$z^2 + 2\beta z - 1 = 0$, so $z = \sqrt{\beta^2 + 1} - \beta$. As $z =
\frac{w^2}{\sqrt{108}}$, and $w = \tilde\rho$ at closest approach, we
deduce that
\be
\sqrt{\beta^2 + 1} - \beta = \frac{2\sqrt{2}}{a} \left( 1 - \sqrt{1 -
    \frac{a^2}{8}} \right) \,,
\ee
which fortunately simplifies to
\be
\beta = \sqrt{\frac{8}{a^2} - 1} \,.
\label{relabeta}
\ee

To understand vortex scattering, we focus on the coordinate $w$.
The great circle segments in the $z$-plane become segments of
quartic curves in the $w$-plane, with reflection symmetry in the
real axis, although we do not need to know these curves in detail.
Because the flat-cone parts of a geodesic are straight (in
the coordinate $w$), scattering only occurs on the curved segment, i.e. on
the spherical cap. The scattering angle depends on the change of direction of
this segment, between its start and end points. The direction of an
infinitesimal part of a segment is the argument of $dw$, and by
reflection symmetry, the change of direction along this segment is
directly related to the difference between the arguments of $dw$ and
$d\bar{w}$, that is, to the argument of $\frac{dw}{d{\bar w}}$, evaluated
at the endpoint of the segment.

\begin{figure}
\center
\includegraphics[width=0.45\columnwidth]{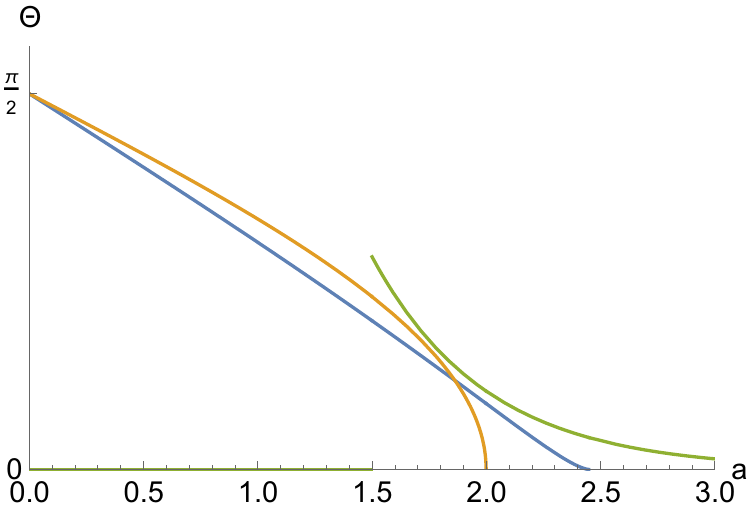}
\includegraphics[width=0.45\columnwidth]{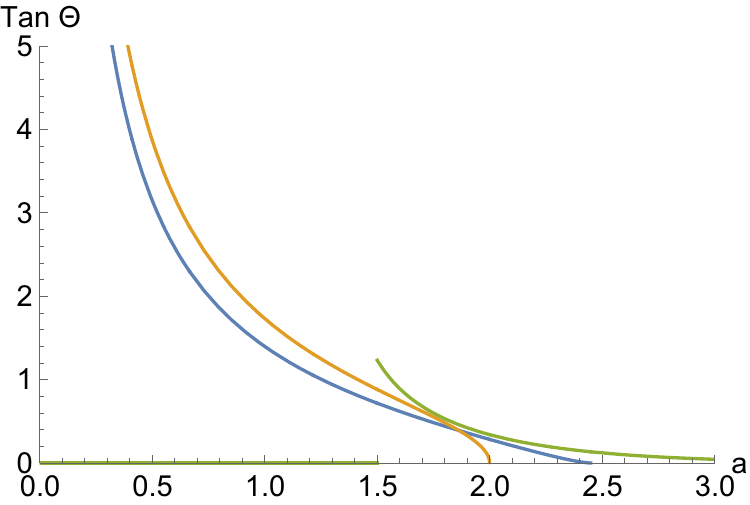}
\caption{The scattering angle $\Theta$, and $\tan\Theta$, as functions
of the impact parameter $a$: the spherical cap
approximation (blue); the flat cap approximation (orange);
the asymptotic approximation (green) \cite{MSp}.}
\label{plot-tanTheta}
\end{figure}

To calculate the argument of $\frac{dw}{d{\bar w}}$, we note that
$w = (108)^{1/4} \sqrt{z}$ so
\be
dw = \frac{(108)^{1/4}dz}{2\sqrt{z}} \quad {\rm and} \quad
d{\bar w} = \frac{(108)^{1/4}d{\bar{z}}}{2\sqrt{\bar z}} \,.
\ee
Therefore
\be
\frac{dw}{d{\bar w}} = \sqrt{\frac{\bar{z}}{z}} \frac{dz}{d{\bar{z}}} \,.
\ee
Next, taking the differential of the equation (\ref{circle}) for a great
circle segment, we find that
\be
\frac{dz}{d{\bar z}} = - \frac{\beta + z}{\beta + \bar{z}}
\ee
so everywhere along the segment,
\be
\frac{dw}{d{\bar w}} = - \sqrt{\frac{\bar{z}}{z}} \,
\frac{\beta + z}{\beta + \bar{z}} \,.
\label{direction}
\ee
The endpoints in the $z$-plane are where the circle (\ref{circle})
intersects the boundary circle of the spherical cap, $|z| =
\frac{1}{\sqrt{3}}$. The endpoint with positive imaginary part is
\be
z = \frac{1}{3\beta} (1 + i \sqrt{3\beta^2 - 1}) \,.
\ee
After some manipulation, we find from (\ref{direction}) the
endpoint value
\be
\arg \frac{dw}{d\bar{w}} = \pi + \arctan{\sqrt{3\beta^2 - 1}} -
2\arctan \left(\half\sqrt{3\beta^2 - 1} \right) \,,
\ee
and simple geometry shows that the scattering angle $\Theta$ along this
geodesic is
\be
\Theta = \pi - \arg \frac{dw}{d\bar{w}}
= 2\arctan \left(\half\sqrt{3\beta^2 - 1} \right)
- \arctan{\sqrt{3\beta^2 - 1}} \,.
\ee
Expressed in terms of the impact parameter $a$, using (\ref{relabeta}), this
becomes
\be
\Theta = 2 \arctan \sqrt{\frac{6}{a^2} - 1} - \arctan
\left( 2\sqrt{\frac{6}{a^2} - 1} \right) \,.
\ee
Finally, using the subtraction and double-angle formulae for the
tangent function, we conclude that
\be
\tan\Theta = \frac{(6-a^2)^{3/2}}{a(9-a^2)} \,.
\ee
This is valid for $a \le \sqrt{6}$, and for larger impact parameters
there is no scattering. As $a \to 0$, the scattering angle approaches
$90^\circ$, the expected result for two vortices in a head-on collision
\cite{Samols}. Fig. \ref{plot-tanTheta} shows the scattering angle as
a function of impact parameter, using this spherical cap approximation to
the 2-vortex moduli space.

\subsection{Flat cap approximation}

For our second, cruder approximation to the metric on $\M$, we attach
a flat cap -- a disc -- to the top of a truncated flat cone, with the join
at $\rho = \rho_0 = 2$. As before, $w = \rho e^{i\theta}$ with $\rho \geq 0$
and $\theta \in [-\half\pi,\half\pi]$. The total metric is
\be
ds^2_{(2)}=f^2_{(2)}(\rho) \left( d\rho^2 + \rho^2d\theta^2 \right)
\ee
where
\be
f^2_{(2)}(\rho) = \left\{
\begin{array}{cc}
\half \pi \rho^2 & \rho \leq 2 \,, \\
& \\
2\pi & \rho \geq 2 \,.
\end{array}
\right.
\ee
This matches the asymptotic, flat-cone metric for $\rho \geq 2$. For
$\rho \le 2$, the change of coordinate $z = \frac{1}{4} w^2 =
\frac{1}{4} \rho^2 e^{i2\theta}$ converts the metric to
$ds^2 = 2\pi dz d\bar{z}$, with $|z| \le 1$, a flat-disc metric.
The missing part of the cone has area $4\pi^2$, whereas the disc has
area $2\pi^2$. The area deficit is again $2\pi^2$, as required.

In this approximation, a geodesic has straight segments on the
cone joined to a straight segment on the flat cap. Intrinsically, the
tangent to the geodesic is continuous, even though the complete
surface has a delta-function curvature at the join. In the $z$-plane,
the geodesic segment on the flat cap is
\be
z + \bar{z} = 2X \,,
\ee
where the parameter $X$ is real and non-negative, and in the range
$0 \le X \le 1$. This segment, parallel to the imaginary
axis, has the reflection symmetry that we imposed earlier.

In the $w$-plane, this geodesic segment becomes
\be
w^2 + {\bar{w}}^2 = 8X \,,
\label{hyperbola}
\ee
which is a rectangular hyperbola. Its closest approach to the origin is
${\tilde\rho} = 2\sqrt{X}$. To find the relation between
$X$ and the impact parameter $a$ we again use the conservation of
$J^2/E$, leading to $f^2_{(2)}({\tilde\rho}) {\tilde\rho}^2 = 2\pi
a^2$, the analogue of (\ref{closest}), which implies that
\be
\half \pi {\tilde\rho}^4 = 2\pi a^2 \,.
\ee
Therefore,
\be
X = \half a \,.
\ee

The scattering angle of the vortices depends only on the change of
direction in the $w$-plane of the flat-cap geodesic segment
between its endpoints, again given by $\frac{dw}{d\bar{w}}$. The
differential of eq.(\ref{hyperbola}) for this segment implies that
\be
\frac{dw}{d\bar{w}} = - \frac{\bar{w}}{w} \,.
\ee
The endpoint (with positive imaginary part) in the $z$-plane is where
$|z| = 1$, so $z = X + i\sqrt{1-X^2}$, which converts to
$w = \sqrt{2} \, ({\sqrt{1+X} + i\sqrt{1-X}})$. Therefore
\be
\arg \left(-\frac{\bar{w}}{w} \right) =
\pi - 2 \arctan \left(\frac{\sqrt{1-X}}{\sqrt{1+X}} \right) \,,
\ee
so the scattering angle is
\be
\Theta = 2 \arctan \left(\frac{\sqrt{1-X}}{\sqrt{1+X}} \right) \,,
\ee
and from the double-angle formula,
\be
\tan\Theta = \frac{\sqrt{1-X^2}}{X} \,.
\ee
As $X = \half a$, the scattering angle as a function of impact
parameter $a$ in the flat cap approximation is
\be
\Theta = \arctan \left( \frac{\sqrt{2-a}}{\sqrt{2+a}} \right) \,,
\quad {\rm so} \quad
\tan\Theta = \frac{\sqrt{4-a^2}}{a} \,,
\label{flatcapscatt}
\ee
both functions being shown in Fig. 1. This is not such a good
approximation for the scattering angle as that obtained using the
spherical cap approximation. In particular, the scattering angle
(\ref{flatcapscatt}) has an unwanted square root singularity as $a \to 2$.

\subsection{Approximations for the mode frequencies}

The squared frequency $\omega_1^2$ of the lowest mode varies with the vortex
separation. It monotonically increases from approximately
$\omega_1^2(0)=0.5378$ to $\omega_1^2(\infty) = 0.7747$ (the squared frequency
of the 1-vortex radial shape mode) as $\rho$ increases from $0$ to
$\infty$\footnote{More precise frequencies, together with numerical
error estimates, are given in ref.\cite{AGMM}. The continuum
spectrum has frequencies $\omega \ge 1$.}. $\omega_1^2(\rho)$ can be
approximated by the following rather simple function, 
\be
\omega_1^2(\rho)=\left\{ 
\begin{array}{lc}
\omega_1^2(0)+\frac{1}{R^3(R+2)}
(\omega_1^2(\infty)-\omega_1^2(0))\rho^4 & \rho \leq R \\
& \\
\omega_1^2(\infty)-\frac{2}{R+2}
(\omega_1^2(\infty)-\omega_1^2(0)) e^{2(R-\rho)} & \rho \geq R \,,
\end{array}
\right.
\label{approxfreq1}
\ee
which is continuous and has continuous first derivative. It can be expressed
in terms of the coordinate $r=\sqrt{x^2+y^2}$ using the relation
$\rho^2=\sqrt{108} \, r$. The form of (\ref{approxfreq1}) is motivated by the
facts that near the origin the squared frequency grows quadratically with
$r$, i.e quartically with $\rho$, and that it approaches $\omega_1^2(\infty)$
exponentially with the vortex separation $2\rho$. Here, $R$ is a
scale parameter, and a good fit is achieved with $R=2$. Because
$\omega_1^2$ increases with the vortex separation, it will generate an
attractive interaction. 

The squared frequency of the second mode can be approximated as
\be
\omega^2_2(\rho)=
\omega^2_2(0) - (\omega^2_2(0)-\omega^2_2(\infty))
\left(1-e^{-0.2 \rho^2} \right),  
\label{secondfreq}
\ee
where $\omega_2^2(0)=0.9747$ is the degenerate frequency of the
second and third modes at coincidence, i.e. at the apex of the moduli
space, and $\omega_2^2(\infty) = \omega_1^2(\infty) = 0.7747$. $\omega_2$
depends linearly on $\rho^2$ near the apex, and continues smoothly to
negative $\rho^2$ to give the frequency $\omega_3$ of the third mode.
(In this context, $r$ is proportional to $|\rho^2|$.) Close to
$\rho^2 = -1$, the third mode hits the continuum threshold $\omega=1$,
and disappears. 

\section{Model for the excited, lowest-frequency mode}

\subsection{Collective coordinate model}

Here we consider a collective coordinate model for 2-vortex dynamics
with the lowest mode excited. For vortices approaching from a large
separation, this mode represents an in-phase superposition of the
radial shape modes on each vortex.

This model is quite simple. Over the 2-vortex moduli
space $\M$ with metric $ds^2 = \Omega(x,y)(dx^2 + dy^2)$, we assume there is
defined a harmonic oscillator with normal coordinate $\eta$ and
position-dependent frequency $\omega_1(x,y)$. The following
Lagrangian couples the excited oscillator to motion through $\M$:
\be
L = \half \Omega(x,y) ({\dot x}^2 + {\dot y}^2) + \half {\dot\eta}^2 - \half
\omega_1^2(x,y) \eta^2 \,.
\label{Lagran}
\ee
There are no cross terms in the kinetic energy, because the moduli
space directions are zero modes of the 2-vortex fields,
whereas the oscillator direction is a positive-frequency shape mode,
and these modes are orthogonal.

The equations of motion derived from $L$ are
\bea
\frac{d}{dt} (\Omega {\dot x}) - \half \pr_x \Omega ({\dot x}^2 + {\dot y}^2)
+ \omega_1 \pr_x \omega_1 \, \eta^2 &=& 0 \,, \label{xeqs}\\
\frac{d}{dt} (\Omega {\dot y}) - \half \pr_y \Omega ({\dot x}^2 + {\dot y}^2)
+ \omega_1 \pr_y \omega_1 \, \eta^2 &=& 0 \,, \label{yeqs} \\
\ddot\eta + \omega_1^2 \eta &=& 0 \,. \label{etaeq}
\eea
Equations (\ref{xeqs})-(\ref{yeqs}) can be expanded out, giving
\bea
\Omega {\ddot x} + \half \pr_x \Omega \, {\dot x}^2 + \pr_y \Omega \, {\dot x}
{\dot y} - \half \pr_x \Omega \, {\dot y}^2
+ \omega_1 \pr_x \omega_1 \, \eta^2 &=& 0 \,, \label{xeqexpand}\\
\Omega {\ddot y} - \half \pr_y \Omega \, {\dot x}^2 + \pr_x \Omega \, {\dot x}
{\dot y} + \half \pr_y \Omega \, {\dot y}^2
+ \omega_1 \pr_y \omega_1 \, \eta^2 &=& 0 \,.
\label{yeqexpand}
\eea
Here, the coefficients of the quadratic terms in velocity (divided by
$\Omega$) encode the Levi-Civita connection on $\M$.

\subsection{Numerical results}

For the numerical analysis of this model we use the spherical
cap approximation to $\Omega$, i.e. to the geometry of ${\cal M}$,
and the approximation (\ref{approxfreq1}) for $\omega_1^2$. Both have
rotational symmetry. For simplicity, we restrict ourselves to head-on
collisions. Thus, it is consistent to put $y\equiv 0$ identically
in (\ref{xeqexpand}) and identify positive $x$ with motion of the vortex pair
along the horizontal axis while negative $x$ corresponds to motion
along the vertical axis. Each passage through $x=0$ corresponds to
$90^\circ$ scattering, and we call this a bounce. For this restricted motion,
\bea
\Omega {\ddot x} + \half \pr_x \Omega \, {\dot x}^2
+ \omega_1 \pr_x \omega_1 \, \eta^2 &=& 0 \,,  \\
\ddot\eta + \omega_1^2 \eta &=& 0 \,. \label{2CCM}
\eea

We assume that at $t=0$ the vortices are well separated along
the horizontal axis and approaching each other. In our
numerics we assumed $x(0)=3$, corresponding to an initial vortex
separation $2\rho \approx 11$, and $v_{\rm in} = -{\dot x}(0) > 0$.
Because the mode-mediated force between the vortices is always attractive
there is at least one collision, i.e. $x(t) = 0$ at least once.
\begin{figure}
\includegraphics[width=0.5\columnwidth]{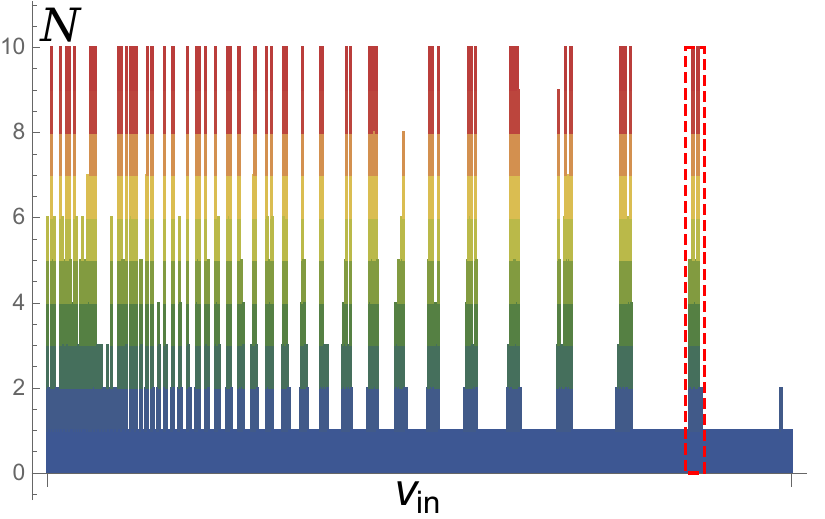}
\includegraphics[width=0.5\columnwidth]{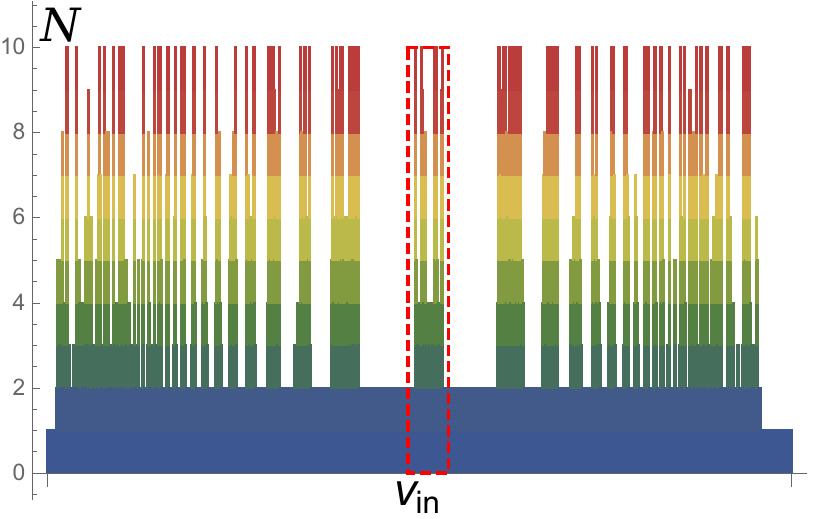}
\includegraphics[width=0.5\columnwidth]{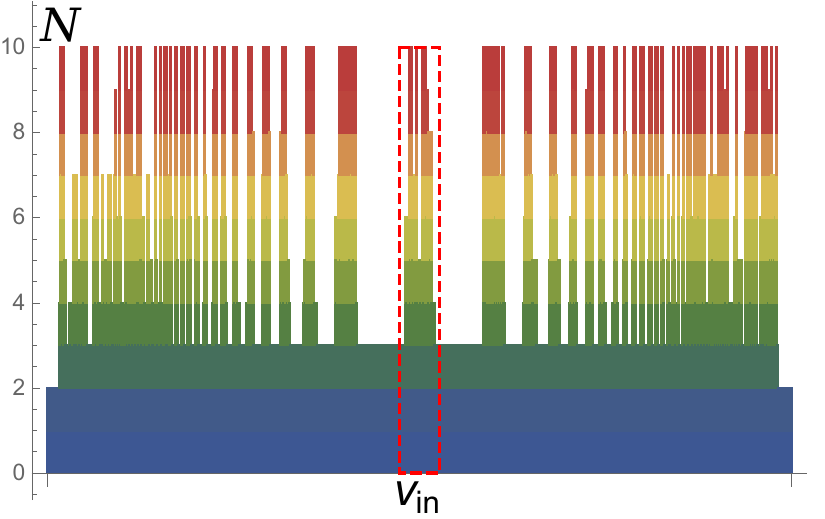}
\includegraphics[width=0.5\columnwidth]{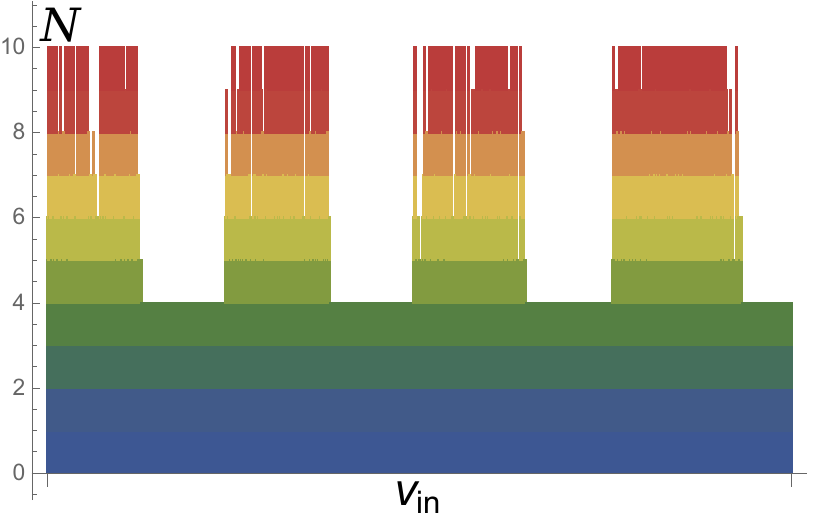}
\caption{Number of bounces $N$ as a function of initial velocity
$v_{\rm in}$. Here $N=10$ denotes 10 or more bounces.
{\it Upper left}: multi-bounces immersed in 1-bounce windows, $v_{\rm in} \in
[0,0.1]$; {\it Upper right}: multi-bounces immersed in 2-bounce
windows, $v_{\rm in} \in [0.0862,0.088]$; {\it Bottom left}:
multi-bounces immersed in 3-bounce windows, $v_{\rm in} \in [0.0870892,
0.0871576]$; {\it Bottom right}: multi-bounces immersed in 4-bounce
windows, $v_{\rm in} \in [0.087122032, 0.0871248364]$.}
\label{window-plot}
\end{figure} 
\begin{figure}
\includegraphics[width=0.5\columnwidth]{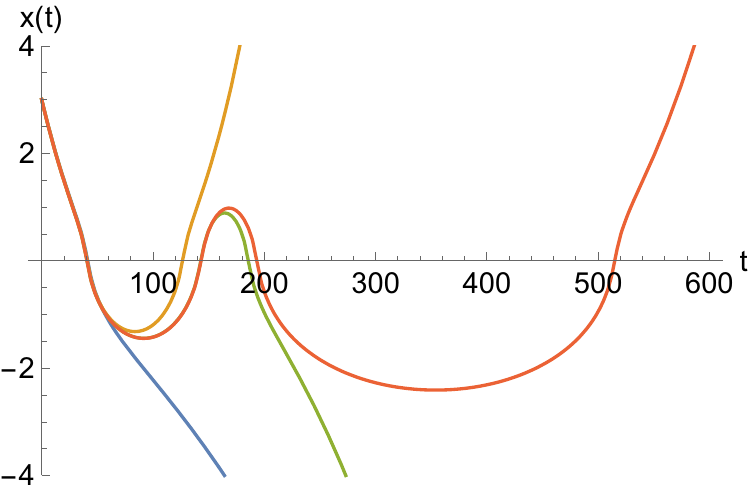}
\includegraphics[width=0.5\columnwidth]{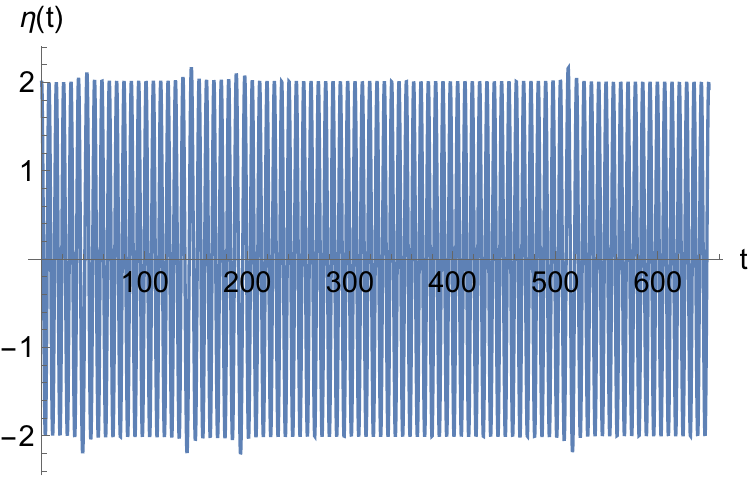}
\caption{{\it Left}: examples of trajectories $x(t)$ of the 2-vortex modulus:
1-bounce, $v_{\rm in}=0.086$; 2-bounce, $v_{\rm in}=0.087$; 3-bounce,
$v_{\rm in}=0.08757$ and 4-bounce, $v_{\rm in}=0.087571687055$. {\it Right}:
evolution of the mode amplitude $\eta(t)$ for the 4-bounce solution.}
\label{trajectory-plot}
\end{figure}
The lowest mode is excited with initial amplitude $\eta(0)=2$ and
$\dot{\eta}(0)=0$. $v_{\rm in}$ is then varied. One should remember
that $v_{\rm in}$ is not the initial velocity of vortices
in the physical plane, although the difference is rather small.
The relation between the variables $\rho$ and $x$ gives
\be
v_{\rm in}^{\rm phys}=\left( \frac{27}{4} \right)^{1/4}
\frac{v_{\rm in}}{\sqrt{x(0)}}=\left( \frac{3}{4} \right)^{1/4}
v_{\rm in} \,.
\ee

Our main result is that we find a {\it chaotic structure} in the
scattering as $v_{\rm in}$ increases. The usual $90^\circ$
scattering (1-bounce) arising from the geodesic approximation is, in a rather
chaotic way, replaced by multi-bounce windows, where colliding
vortices form a quasi-bound state performing $N$ bounces --
each being a $90^\circ$ scattering. For sufficiently large amplitude,
the interchanging sequence of 1-bounce and multi-bounce windows
starts from arbitrary small initial velocity and ends when $v_{\rm in}$
exceeds a critical velocity $v_{\rm cr}$. For $\eta(0) = 2$ we find that
$v_{\rm cr}=0.0988$, and for $v_{\rm in} > v_{\rm cr}$ we observe only
single bounces, see Fig. \ref{window-plot}, upper left. The figure
shows the number of bounces (from 1 to $N \geq 10$) for initial
velocities in the range $v_{\rm in}\in [0,0.1]$.

\begin{figure}
\includegraphics[width=0.5\columnwidth]{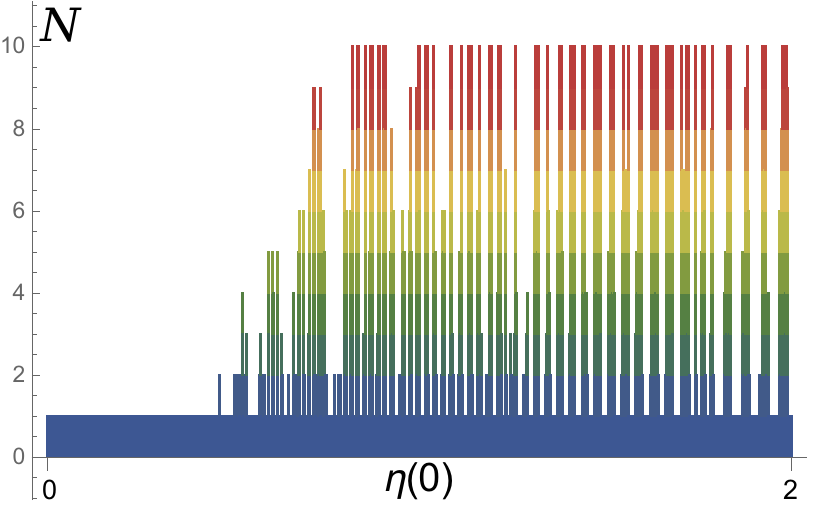}
\includegraphics[width=0.5\columnwidth]{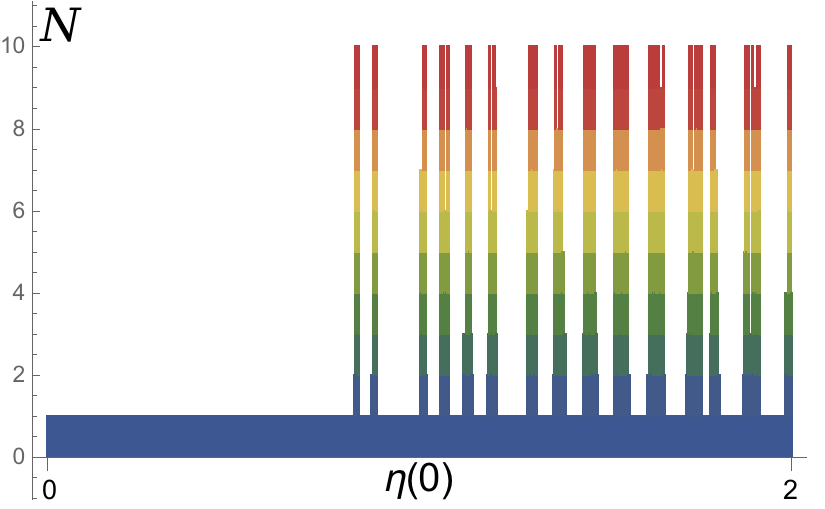}
\caption{Number of bounces as a function of initial amplitude of the
mode in the range $\eta(0) \in [0,2]$. {\it Left}: $v_{\rm in}=0.005$;
{\it Right}: $v_{\rm in}=0.015$.}
\label{amplitude_change-plot}
\end{figure}
 
This chaotic structure has an approximately self-similar pattern.
In Fig. \ref{window-plot}, upper right, we show the $N \geq 3$
bounces immersed among 2-bounce scatterings, for $v_{\rm in} \in
[0.0862,0.088]$. This structure repeats. The lower panels of
Fig. \ref{window-plot} show $N \geq 4$ bounces immersed among
3-bounce scatterings, for $v_{\rm in}\in [0.0870892, 0.0871576]$
and $N \geq 5$ bounces immersed among 4-bounce scatterings,
for $v_{\rm in} \in[0.087122032, 0.0871248364]$.
 
In Fig. \ref{trajectory-plot}, left, we show examples of 1-, 2-, 3-
and 4-bounce solutions $x(t)$ for $v_{\rm in}=0.086, 0.087, 0.08757$ and
$0.087571687055$ respectively. Clearly, a tiny change in the initial
conditions can lead to a dramatic change in the scattering. In
Fig. \ref{trajectory-plot}, right, we plot the time evolution
$\eta(t)$ of the mode amplitude for this 4-bounce solution. The maximum
amplitude is almost constant but briefly grows during the collisions.

\begin{figure}
\includegraphics[width=0.5\columnwidth]{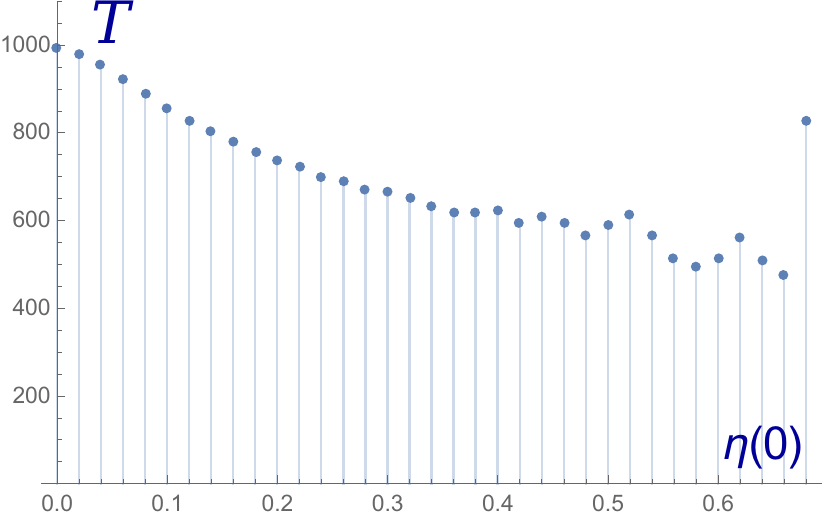}
\includegraphics[width=0.5\columnwidth]{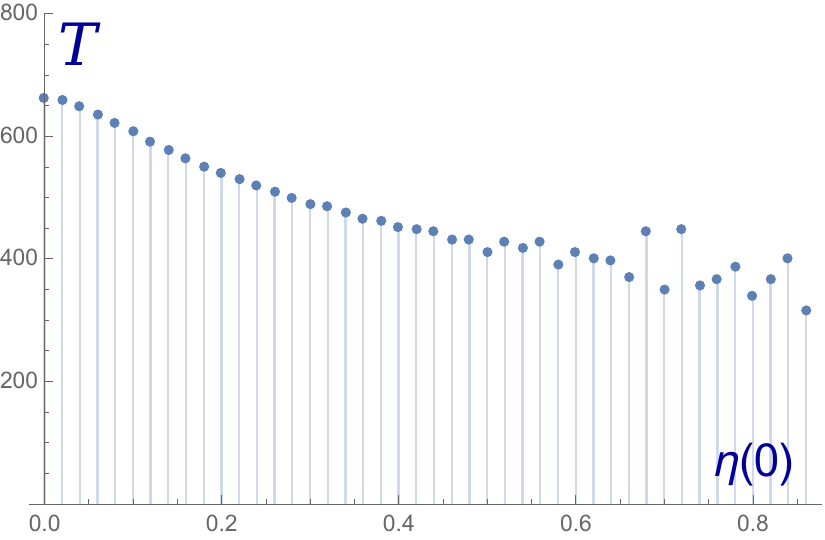}
\caption{Collision time $T$ as a function of the initial amplitude
of the mode $\eta(0)$. $T$ is the time for the trajectory to reach
$x=-3$, starting from $x=3$. {\it Left}: $v_{\rm in}=0.010$;
{\it Right}: $v_{\rm in}=0.015$.}
\label{time-acceler}
\end{figure}

In Fig. \ref{amplitude_change-plot} we show how the structure of
bounces changes if we vary the initial mode amplitude $\eta(0)$ but
fix the initial velocity $v_{\rm in}$. For example, for
$v_{\rm in}=0.05$ the first multi-bounce occurs when $\eta(0)=0.504$
whereas for $v_{\rm in}=0.15$ it occurs when $\eta(0)=0.826$. For sufficiently
small values of the amplitude there is
always a regime with one bounce, i.e. a single $90^\circ$ scattering.
In this {\it quasi-geodesic} regime the geodesic dynamics is only softly
modified by the mode excitation, making the vortex-vortex collision
faster; see Fig. \ref{time-acceler}, where we plot the time $T$ at which the
trajectory reaches $x=-3$, starting from $x=3$. $T$ decreases as the
initial amplitude of the mode grows, a consequence of the attractive
force triggered by the non-zero mode amplitude. However, above a
critical value of $\eta(0)$ the chaotic multi-bounce behaviour starts.
At this point, $T$ jumps and the geodesic approximation breaks down
completely. Not surprisingly, the quasi-geodesic regime is larger if
$v_{\rm in}$ is larger. In the next subsection we will analyze this
regime from an adiabatic point of view.

We remark that although the positions of the 1-bounce and multi-bounce
collisions are quite sensitive to details of our collective model,
that is, to changes of $\omega_1$ and $\Omega$, the
chaotic multi-bounce structure is robust. Therefore we expect
that vortex scattering in the full field theory with the lowest mode
excited will exhibit the same features. In fact, preliminary
results from numerical simulations of the field theory confirm
the validity of our collective model \cite{Rees}.

All these results resemble what is observed in kink-antikink scattering
in $\phi^4$ theory in (1+1) dimensions. There also, there is a
chaotic sequence of multi-bounce windows and annihilation regions
called bion chimneys \cite{sug, CSW}, with a strong dependence
on the initial relative velocity of the kinks, and on the initial
shape mode amplitude. In particular, the accumulation of 2-vortex
multi-bounces as $v_{\rm in}$ tends to 0 (for fixed initial mode
amplitude) possesses its counterpart in scatterings between a
wobbling kink and antikink \cite{AI}. Such a chaotic pattern is
explained in terms of a resonant energy transfer mechanism between the
kink kinetic energy and shape mode energy \cite{sug, CSW}, and it has
recently been established that a collective model with two degrees of
freedom explains the observed dynamics well \cite{MORW}. The main
difference from vortices is that the chaotic behaviour in
kink-antikink collisions occurs even without an initial excitation
of the shape mode, because there is automatically an attractive
force between kink and antikink. For the BPS 2-vortices, an
excitation of the lowest shape mode is needed to generate an attraction. 

\subsection{Adiabatic approximation}

We now treat the oscillator as a fast variable, and the motion
through $\M$ as slow. It is also important that the mode
amplitude is relatively small. More precisely, we assume that
$\omega_1$ and $\Omega$, together with their $x$- and $y$-derivatives,
are $O(1)$, and that ${\dot x}$ and ${\dot y}$ are $O(\eps)$, with
$\eps$ small. We wish ${\ddot x}$ and ${\ddot y}$ to be $O(\eps^2)$,
and see from eqs.(\ref{xeqexpand}) and (\ref{yeqexpand}) that
the oscillator amplitude $\eta$ needs to be $O(\eps)$. With these
assumptions, the solution of eq.(\ref{etaeq}) for the oscillator,
in the adiabatic approximation, takes the form
\be
\eta(t) = A(x(t),y(t)) \, \cos \left( \int_0^t \omega_1(x(t'),y(t'))
\, dt' \right) \,,
\label{etaansatz}
\ee
where we have chosen the time origin to coincide with an
instantaneous maximal amplitude of $\eta$. The integral is along a path
through $\M$ that is still to be determined using
the equations for $x$ and $y$. The amplitude $A$ is also yet to be
determined, but only depends on the position in $\M$ at
time $t$, and not on the path through $\M$. Clearly,
$A$ needs to be $O(\eps)$.

Differentiating (\ref{etaansatz}) twice with respect to time, we find
that
\be
\ddot\eta + \omega_1^2 \eta =
{\ddot A} \cos  \left( \int_0^t \omega_1(x(t'),y(t')) \, dt' \right)
- \left(2\omega_1{\dot A} + A {\dot\omega_1}\right)
\sin  \left( \int_0^t \omega_1(x(t'),y(t')) \, dt' \right) \,,
\ee
where $\dot\omega_1$ is the total time-derivative of $\omega_1$ along the
path. The first term on the right-hand side is $O(\eps^3)$, and can be
neglected relative to the remaining terms, which are $O(\eps^2)$.
Equation (\ref{etaeq}) for the oscillator is therefore satisfied provided that
\be
2\omega_1{\dot A} + A {\dot\omega_1} = 0 \,.
\ee
The solution is
\be
A = \frac{C}{\sqrt{\omega_1}}
\ee
with $C$ a constant, so $A$ is indeed like $\omega_1$, being just a function
of the instantaneous position $(x(t),y(t))$. $C$ is the adiabatic
invariant of the oscillator, remaining constant despite the oscillator
having a varying frequency $\omega_1$ along the path in $\M$. For
consistency, $C$ is $O(\eps)$.

We now derive the reduced, adiabatic equations of motion for $x$
and $y$, using the approximate solution for the oscillator
\be
\eta(t) = \frac{C}{\sqrt{\omega_1(x(t),y(t))}} \,
\cos \left( \int_0^t \omega_1(x(t'),y(t')) \, dt' \right) \,.
\label{etasoln}
\ee
We need to take the time-average of $\eta^2$ over one period of the
oscillator, to derive the average force that acts. This is
$\langle \eta^2 \rangle = \half A^2 = \half \frac{C^2}{\omega_1}$. The reduced
equations are therefore eqs.(\ref{xeqexpand}) and (\ref{yeqexpand})
with $\omega_1 \eta^2$ replaced by $\half C^2$. More convenient is to
give the reduced Lagrangian, from which these equations follow, namely
\be
L_{\rm red} = \half \Omega(x,y) ({\dot x}^2 + {\dot y}^2) - \half C^2
\omega_1(x,y) \,.
\label{Lagranred}
\ee
The constant $C^2$ is determined by the oscillator's initial
conditions, and implicitly, ${\dot x}^2, {\dot y}^2$ and $C^2$ are
all $O(\eps^2)$.

In summary, the oscillator dynamics generates a potential energy
$\half C^2 \omega_1(x,y)$ that, together with the conformal factor $\Omega$,
governs adiabatic motion through $\M$. This motion has
the conserved energy
\be
E_{\rm red} = \half \Omega(x,y) ({\dot x}^2 + {\dot y}^2) + \half C^2
\omega_1(x,y) \,,
\ee
which matches the conserved energy for the original Lagrangian (\ref{Lagran})
if we use the time-averaged oscillator energy. The latter is
\be
\langle E_{\rm osc} \rangle = \half A^2 \omega_1^2 = \half C^2 \omega_1 \,,
\ee
where we have ignored the contribution of $\dot A$ relative to that of
$A$. Note that the adiabatic invariant $C^2$ is
$\frac{2}{\omega_1} \langle E_{\rm osc} \rangle$.

The discussion so far has needed to assume no symmetry property for
$\omega_1$ and $\Omega$. However, in the context of the 2-vortex
moduli space ${\cal M}$ -- the
rounded cone -- both the frequency $\omega_1$ of the lowest shape mode
and the metric conformal factor $\Omega$ have $O(2)$ rotational
symmetry. It is therefore convenient to use the polar coordinates
$x = r\cos\varphi, \, y = r\sin\varphi$ on ${\cal M}$, and the radial functions
$\omega_1(r)$ and $\Omega(r)$. $r=0$ is the apex of $\M$, where
the vortices are coincident, and $r$ extends to $+\infty$ as the
vortices separate. $\omega(r)$ and $\Omega(r)$ are positive and have
zero derivative at $r=0$. The angle $\varphi$ has the range $\varphi
\in [-\pi,\pi]$.

The Lagrangian (\ref{Lagran}) for the shape mode dynamics coupled
to the moduli space motion simplifies as a result of the
rotational symmetry, and in particular, the reduced Lagrangian
(\ref{Lagranred}) simplifies to
\be
L_{\rm red} = \half \Omega(r) ({\dot r}^2 + r^2 {\dot \varphi}^2) -
\half C^2 \omega_1(r) \,.
\ee
There are two constants of motion for $L_{\rm red}$, the energy and
angular momentum
\bea
E &=& \half \Omega(r) ({\dot r}^2 + r^2 {\dot \varphi}^2) +
\half C^2 \omega_1(r) \,, \\
J &=& \Omega(r) r^2 {\dot \varphi} \,.
\eea
After eliminating ${\dot\varphi}$ from the energy in favour of the
angular momentum, the radial motion can be found by quadrature,
provided we know both $\omega_1(r)$ and $\Omega(r)$, and all the initial data,
including that for the oscillator.

As $\omega_1(r)$ is an increasing function of $r$, the potential is
attractive, but the effective potential in the reduced dynamics,
\be
V_{\rm red}=\frac{1}{2} C^2\omega_1(r) + \frac{J^2}{2 \Omega(r)r^2} \,, 
\ee
includes a centrifugal term, so can be can be attractive or repulsive.
There can therefore be both bounded and scattering solutions for the
adiabatic 2-vortex dynamics when the lowest shape mode is excited.
This contrasts with the dynamics in the absence
of shape mode oscillations, where the motion on ${\cal M}$ follows
geodesics, and consists purely of scattering trajectories \cite{Samols}.
The existence of bound orbits depends on the ratio between the
adiabatic invariant and the angular momentum, $C/J$.

The simplest motion, using this adiabatic approximation, is a head-on
collision, with the vortices approaching from a large separation at
some finite velocity. The timing for this is shown in Fig. \ref{time-acceler}.

\section{Model for higher-frequency shape modes}

\subsection{Collective coordinate model}

Recall that the higher-frequency shape modes (the second and third modes)
become degenerate at the apex of the moduli space $\M$, and that the
third mode enters the continuum
close to the apex. A model for the higher-frequency modes needs to
allow for their interaction, as their frequencies are close together,
but it needs to be constructed only in the inner region of $\M$, around
the apex. Here we can approximate the geometry of $\M$ by a flat cap
(the spherical cap approximation would be a refinement), and the mode
frequencies can be approximated as having a linear dependence on
$|z|$, where $z = x+iy = r e^{i\varphi}$ is the suitably
normalized complex coordinate centred at the apex of $\M$.

The coordinate $z$ is a parametrization for the 2-vortex
gauge and Higgs fields over $\M$; similarly the vector space of
higher-frequency shape modes and their amplitudes, involving
deformations of the gauge and Higgs fields and a background gauge
condition, can be parametrized by an abstract pair of amplitudes $\zeta$ and
$\chi$. These are initially complex, but we will impose a reality
condition below. There is $O(2)$ rotational symmetry about $z=0$, and
the modes transform at $z=0$ as a doublet of $O(2)$.

The potential energy for the modes is
constructed using the $2 \times 2$ hermitian matrix
\be
M = \begin{pmatrix} \lambda & \alpha(x-iy) \\ \alpha(x+iy) &
\lambda \end{pmatrix}
= \begin{pmatrix} \lambda & \alpha z^* \\ \alpha z & \lambda \end{pmatrix} \,,
\ee
where $\lambda$ and $\alpha$ are real and positive constants; $\lambda$ is the
degenerate eigenvalue of $M$ when $z=0$. The eigenvalues of $M$ split
for $z \ne 0$, becoming $\lambda \pm \alpha |z|$. The complete Lagrangian of
the model, with standard kinetic terms and a quadratic potential obtained
using $M$, is
\bea
L &=& \half \left\{ {\dot z}^* {\dot z} + {\dot \zeta}^*{\dot \zeta}
  + {\dot \chi}^*{\dot \chi} - (\zeta^* \ \chi^*)
\begin{pmatrix} \lambda & \alpha z^* \\ \alpha z & \lambda \end{pmatrix}
\begin{pmatrix} \zeta \\ \chi \end{pmatrix} \right\} \nonumber \\
&=& \half \left\{ {\dot r}^2 + r^2 {\dot\varphi}^2 + {\dot \zeta}^*{\dot \zeta}
  + {\dot \chi}^*{\dot \chi} - (\zeta^* \ \chi^*)
\begin{pmatrix} \lambda & \alpha r e^{-i\varphi} \\ \alpha r e^{i\varphi}
  & \lambda \end{pmatrix}
\begin{pmatrix} \zeta \\ \chi \end{pmatrix} \right\} \,.
\label{fullLag}
\eea
$\zeta$ and $\chi$ are the (undiagonalised) amplitudes of the modes
whose frequencies are controlled by the $z$-dependent matrix $M$. Before
looking at the equations of motion it helps to say more about the
eigenvectors of $M$.

A generic $2 \times 2$ hermitian matrix is of the form $M = \lambda +
{\bf a} \cdot {\boldsymbol\sigma} = \lambda + a_1\sigma_1 +
a_2\sigma_2 + a_3\sigma_3$, where $\sigma_1 \,, \sigma_2 \,, \sigma_3$
are the Pauli matrices. This has eigenvalues $\lambda \pm
|{\bf a}|$, so the eigenvalues are degenerate only when ${\bf a} =
0$. Generally, in a family of $2 \times 2$ hermitian matrices,
the eigenvalues degenerate on a submanifold of real codimension 3 in the
parameter space. But for our problem we have only two real moduli, and
degeneracy still occurs. The reason is that $M$ is
a family of hermitian matrices with a ``real'' structure. If $a_2$
were zero, then $M$ would be manifestly real, and we could seek real
eigenvectors. Degeneracy would then occur when $a_1 = a_3 = 0$,
a codimension-2 condition. It is more convenient in our model to set
$a_3 = 0$, as this simplifies the eigenvectors, but there
is still a ``real'' structure, and eigenvalue degeneracy occurs at
$z=0$, a single point in the 2-dimensional moduli space. The real
structure occurs, because for the vortices in the abelian Higgs model, the
shape mode eigenfunctions are derived from eigenfunctions of a
scalar Schr\"odinger operator with a real potential \cite{AGMM}.

More concretely, the model Lagrangian (\ref{fullLag}) is invariant under 
$\zeta^* \leftrightarrow \chi$ and we can impose the ``reality'' condition
$\zeta^* = \chi$. This gives a consistent truncation of the equations of
motion. Moreover, it is consistent to impose this condition on the
Lagrangian. We therefore restart from the Lagrangian (\ref{fullLag}),
with $\zeta^*$ replaced by $\chi$ (and $\zeta$ replaced by $\chi^*$),
\be
L = \half {\dot z}^* {\dot z} + {\dot \chi}^*{\dot \chi}
- \lambda \chi^* \chi
- \half \alpha z^* \chi^2 - \half \alpha z \chi^{* \, 2} \,.
\label{realLag}
\ee
The equations of motion now simplify to
\bea
{\ddot z} + \alpha \chi^2 &=& 0 \label{realeq1} \,, \\
{\ddot \chi} + \lambda \chi + \alpha z \chi^* &=& 0 \,, \label{realeq2}
\eea
and there is a conserved energy
\be
E = \half {\dot z}^* {\dot z} + {\dot \chi}^*{\dot \chi}
+ \lambda \chi^* \chi
+ \half \alpha z^* \chi^2 + \half \alpha z \chi^{* \, 2} \,.
\label{ConservE}
\ee
Equation (\ref{realeq1}) implies that the mode oscillations affect
the moduli space motion, and eq.(\ref{realeq2}) implies that the
mode oscillation frequencies at each instant depend on the location $z$
in moduli space. These coupled equations can probably not be solved
analytically, but only numerically.

\subsection{Adiabatic approximation}

However, we can treat the dynamics adiabatically, assuming that
$z$ varies on a timescale much longer than the inverse of the
oscillation frequencies $\sqrt{\lambda \pm \alpha |z|}$. This is just
a little more sophisticated than the adiabatic treatment of the
lowest-frequency oscillation mode, in section 3. The scaling that makes
an adiabatic analysis possible is to suppose
that $z$, $\lambda$ and $\alpha$ are $O(1)$, with $\lambda \pm \alpha |z|$
remaining bounded away from zero, and that ${\dot z} =
O(\varepsilon)$ and ${\ddot z} = O(\varepsilon^2)$, with $\varepsilon$ small.
This requires the mode amplitudes $|\zeta|$ and $|\chi|$ to be
$O(\varepsilon)$, i.e. small, but their oscillation frequencies are $O(1)$.

To proceed, we need a basis of eigenvectors of the matrix $M$ appearing in
eq.(\ref{fullLag}). The eigenvectors/eigenvalues are given by
\be
\begin{pmatrix}
\lambda & \alpha r e^{-i\varphi} \\ \alpha r e^{i\varphi} &
\lambda \end{pmatrix}
\begin{pmatrix} \zeta \\ \chi \end{pmatrix}
= \nu \begin{pmatrix} \zeta \\ \chi \end{pmatrix} \,,
\ee
so the eigenvalues are $\nu_+ = \lambda + \alpha r$ and $\nu_- = \lambda -
\alpha r$, with respective eigenvectors
\be
\begin{pmatrix} \zeta \\ \chi \end{pmatrix}
= \begin{pmatrix} 1 \\ e^{i\varphi} \end{pmatrix}
\quad {\rm and} \quad
\begin{pmatrix} \zeta \\ \chi \end{pmatrix}
= \begin{pmatrix} 1 \\ -e^{i\varphi} \end{pmatrix}
\ee
of squared norm 2. However, these eigenvectors don't satisfy the reality
condition $\zeta^* = \chi$ until we multiply by a suitable phase factor
(which doesn't affect their orthonormality). The ``real'' eigenvectors are
\be
V_+(\varphi) = \pm \begin{pmatrix} e^{-\half i\varphi} \\ e^{\half i\varphi}
\end{pmatrix} \quad {\rm and} \quad
V_-(\varphi) = \pm \begin{pmatrix} ie^{-\half i\varphi} \\ -ie^{\half i\varphi}
\end{pmatrix} \,,
\ee
where the phases are fixed, but there remains an ambiguity in
sign. Note that when $\varphi$ increases by $2\pi$, the sign of each of these
eigenvectors reverses. For a given $r \ne 0$, the eigenspaces are
therefore M\"obius bundles over the circle parametrised by
$\varphi$. In fact, these bundles extend to the entire
punctured plane $r \ne 0$.

It is significant that these M\"obius bundles join up smoothly at
the origin, where the eigenvalues degenerate. There is continuity
of the eigenvectors and eigenvalues along any line in the $z$-plane
that passes through the origin. Consider, for example, the line
$y=0$ with $x$ running from a positive to a negative value. There
is a constant eigenvector $(1 , 1)^{\rm T}$ along this line with
eigenvalue $\lambda + \alpha x$, and another constant
eigenvector $(i , -i)^{\rm T}$ with eigenvalue
$\lambda - \alpha x$. To check this, one has to identify a point with
negative $x$ as having $r = |x|$ and $\varphi = \pi$. Along this line
the lower eigenvalue runs smoothly into the upper eigenvalue, and
vice versa. Such eigenvalue crossing naturally occurs, because the
eigenvalue spectrum exhibits a conical structure over a neighbourhood
of the origin.

We next express the dynamical mode amplitudes in terms of the eigenvectors as
\be
\begin{pmatrix} \zeta \\ \chi \end{pmatrix}
= a_+(t) V_+(\varphi(t)) + a_-(t) V_-(\varphi(t)) \,,
\ee
where $a_+$ and $a_-$ are real. (An arbitrary initial sign
choice for the eigenvectors is made.) The time-derivative is
\be
\frac{d}{dt} \begin{pmatrix} \zeta \\ \chi \end{pmatrix}
= {\dot a}_+ V_+ + a_+ \pr_\varphi V_+ \, {\dot \varphi}
+ {\dot a}_- V_- + a_- \pr_\varphi V_- \, {\dot \varphi} \,.
\ee
The eigenvectors have the simple $\varphi$-derivatives
\be
\pr_\varphi V_+ = -\half V_- \quad {\rm and} \quad
\pr_\varphi V_- = \half V_+ \,,
\ee
so
\be
\frac{d}{dt} \begin{pmatrix} \zeta \\ \chi \end{pmatrix}
= \left({\dot a}_+ + \half a_- {\dot \varphi} \right) V_+
+ \left({\dot a}_- - \half a_+ {\dot \varphi} \right) V_- \,.
\ee
The Lagrangian (\ref{fullLag}) therefore takes the form, in terms of the real
amplitudes $a_+$ and $a_-$,
\be
L = \half\left( {\dot r}^2 + r^2 {\dot\varphi}^2 \right)
+ \left({\dot a}_+ + \half a_- {\dot \varphi} \right)^2
+ \left({\dot a}_- - \half a_+ {\dot \varphi} \right)^2
- (\lambda + \alpha r)a_+^2 - (\lambda - \alpha r)a_-^2 \,,
\ee
and the corresponding equations of motion are
\bea
{\ddot r} - r {\dot \varphi}^2 + \alpha (a_+^2 - a_-^2) = 0 \,,
\label{upper1} \\
r^2 {\dot \varphi} + {\dot a}_+ a_- - {\dot a}_- a_+
+ \half(a_+^2 + a_-^2){\dot \varphi} = \ell \,, \label{upper2} \\
\frac{d}{dt}\left({\dot a}_+ + \half a_- {\dot \varphi} \right)
+ \half\left({\dot a}_- - \half a_+ {\dot \varphi} \right){\dot \varphi}
+ (\lambda + \alpha r)a_+ = 0 \,, \label{upper3} \\
\frac{d}{dt}\left({\dot a}_- - \half a_+ {\dot \varphi} \right)
- \half\left({\dot a}_+ + \half a_- {\dot \varphi} \right){\dot \varphi}
+ (\lambda - \alpha r)a_- = 0 \,. \label{upper4}
\eea
Equation (\ref{upper2}), for $\varphi$, has been integrated
once, and $\ell$ is the conserved angular momentum.

The equations so far are all exact for the model Lagrangian
(\ref{fullLag}), but now we make the adiabatic approximation.
We assume that $r, \varphi$ are $O(1)$, ${\dot r}, {\dot \varphi}$
are $O(\varepsilon)$ and ${\ddot r}, {\ddot \varphi}$ are $O(\varepsilon^2)$,
where $\varepsilon$ is small. For eq.(\ref{upper1}) to be consistent, the
oscillator amplitudes $a_+$ and $a_-$ need to be $O(\varepsilon)$ or
smaller. The following discussion is rather schematic, as the detailed
formulae are not very illuminating.

The basic solution of eqs.(\ref{upper1}) and (\ref{upper2}),
ignoring the oscillator contribution, is a straight-line motion. Let's
orient this line so that $x=b$ and $y = vt$ in Cartesians, where $b$ is
a positive $O(1)$ constant and the velocity $v$ is $O(\varepsilon)$. Then
$\ell = bv$, so
\be
r(t) = \sqrt{b^2 + v^2 t^2} \quad {\rm and} \quad
{\dot \varphi} = \frac{bv}{b^2 + v^2 t^2} \,.
\ee
The straight line needs to
miss the origin for ${\dot \varphi}$ to remain bounded. (We shall consider
a head-on collision below, using the Cartesian coordinate $x$.) Using the
slowly time-dependent $r(t)$ in the oscillator equations
(\ref{upper3}) and (\ref{upper4}), and ignoring all the subleading
terms that depend on ${\dot \varphi}$, we deduce that the adiabatic
solutions for the oscillator amplitudes are
\bea
a_+(t) &=& \frac{C_+}{(\lambda + \alpha r(t))^{\frac{1}{4}}}
\cos \left(\int_0^t \sqrt{\lambda + \alpha r(t')} \, dt' + \gamma_+
\right) \,, \nonumber \\
a_-(t) &=& \frac{C_-}{(\lambda - \alpha r(t))^{\frac{1}{4}}}
\cos \left(\int_0^t \sqrt{\lambda - \alpha r(t')} \, dt' + \gamma_-
\right) \,.
\eea
The constants $C_+$ and $C_-$, together with the phases $\gamma_+$ and
$\gamma_-$, are adiabatic invariants that depend on the initial data.

We can now deduce the modification, due to the mode oscillations, of
the straight-line motion. In eq.(\ref{upper1}) we substitute the
time-averaged values $\langle a_+^2 \rangle =
\frac{C_+^2}{2\sqrt{\lambda + \alpha r(t)}}$ and $\langle a_-^2 \rangle =
\frac{C_-^2}{2\sqrt{\lambda - \alpha r(t)}}$. This results in a
modified radial acceleration that can be attributed to a radial potential
\be
V_{\rm rad} = C_+^2 \sqrt{\lambda + \alpha r}
+ C_-^2 \sqrt{\lambda - \alpha r} + {\rm const.}
\ee
This potential is repulsive for the second ($a_-$) oscillator, as its
frequency increases approaching $r=0$, and attractive for the third
oscillator. In eq.(\ref{upper1}), the term ${\dot a}_+ a_- - {\dot a}_- a_+$
is oscillatory and has zero average, so we ignore it. However, the
time-averaged value of $a_+^2 + a_-^2$ is
$\frac{C_+^2}{2\sqrt{\lambda + \alpha r(t)}} +
\frac{C_-^2}{2\sqrt{\lambda - \alpha r(t)}}$,
and from eq.(\ref{upper2}) we can find the $O(\varepsilon^2)$
modification to ${\dot \varphi}$ due to the mode oscillations.

The analysis has not yet led to any mixing of the oscillation
modes. Mixing occurs if we retain the leading mode-coupling terms in
(\ref{upper3}) and (\ref{upper4}), giving
\bea
{\ddot a_+} + (\lambda + \alpha r) a_+ &=& -{\dot a_-}{\dot \varphi} \,, \\
{\ddot a_-} + (\lambda - \alpha r) a_- &=& {\dot a_+}{\dot \varphi} \,.
\eea
Using the $O(\varepsilon)$ solutions for the oscillator amplitudes and for
${\dot \varphi}$ on the right-hand side, the particular integrals
give $O(\varepsilon^2)$ corrections to the previously determined
$O(\varepsilon)$ homogeneous solutions. In this calculation, it is
sufficient to regard the oscillator frequencies as unvarying.
There is no resonance, because the frequency of each oscillator
differs from the frequency of its forcing term by
$\pm 2r$, and we are assuming $r$ remains $O(1)$.

A special case is if the second mode ($a_-$) is initially excited, but
the third mode is not. This can occur if the vortices approach from
a large separation, where the third mode has disappeared into the
continuum. The analysis above goes through, and the third mode
essentially does not contribute. There is a small excitation of the
third mode due to the forcing by the second mode, if the collision is
not head-on, but the amplitude generated is $O(\varepsilon^2)$.

However, the third mode is much more strongly excited in a head-on
collision, and this is the most interesting case. Let us suppose
the initial motion is along the $x$-axis (the horizontal axis)
in the moduli space, approaching from $+\infty$, and that the second mode
is excited. By symmetry, the motion remains on the $x$-axis and can pass
through $x=0$, leading to $90^\circ$ scattering of the vortices,
although the repulsive potential generated adiabatically by the
mode oscillation may prevent this. Now recall that because of the
conical structure of the oscillator spectrum, it is purely the third
mode that becomes excited if $x$ becomes negative.
The frequency of the excited oscillator, for either sign of
$x$, is the smoothly-varying function $\lambda - \alpha x(t)$,
provided $|x|$ is not large, rather than $\lambda - \alpha r(t)$
(with $r = |x|$).

The relevant equations of motion along the $x$-axis are
\bea
{\ddot x} - \alpha u^2 &=& 0 \label{xeq} \,, \\
{\ddot u} + (\lambda - \alpha x) u &=& 0 \,, \label{ueq}
\eea
obtained from eqs.(\ref{realeq1}) and (\ref{realeq2}) by setting
$z = x$ and $\chi = -iu$, where $x$ and $u$ are real.
For this reduced system, there is a conserved energy
\be
E = \half {\dot x}^2 + {\dot u}^2 + (\lambda - \alpha x) u^2 \,.
\label{RedConservE}
\ee
Again, this is a consistent truncation, and an adiabatic treatment
is feasible if we assume the previous scaling with $\varepsilon$. $u$
oscillates with the slowly-varying frequency $\sqrt{\lambda - \alpha x(t)}$.
The adiabatic solution of (\ref{ueq}) is then
\be
u(t) = \frac{C}{(\lambda - \alpha x(t))^{\frac{1}{4}}} \,
\cos \left( \int_0^t \sqrt{\lambda - \alpha x(t')} \, dt' \right) \,,
\ee
where the constant $C$ is $O(\varepsilon)$. After time-averaging
the oscillatory driving force $u^2$, eq.(\ref{xeq}) becomes
\be
{\ddot x} - \frac{\alpha}{2} \frac{C^2}{\sqrt{\lambda - \alpha x}} = 0 \,,
\ee
with first integral
\be
\half {\dot x}^2 + C^2 \sqrt{\lambda - \alpha x} = E \,.
\label{firstint}
\ee
The constant $E$ can be identified with the total conserved energy.
This is $O(\varepsilon^2)$ and has comparable contributions from the
motion in moduli space and from the oscillating
mode. The oscillating mode contributes an effective potential proportional to
$\sqrt{\lambda - \alpha x}$ to the moduli space dynamics, modifying what
would otherwise be geodesic dynamics with ${\dot x}$ constant. Recall
that $\lambda$ is the degenerate value of the oscillator frequency at
vortex coincidence.

Equation (\ref{firstint}) can be integrated once more to give
\be
t = \int \frac{dx}{\sqrt{2E - 2C^2\sqrt{\lambda - \alpha x}}} \,.
\ee
The integral here is elementary. In terms of the spatial variable
\be
q = \sqrt{1 - \frac{C^2}{E} \sqrt{\lambda - \alpha x}}
\ee
we find the implicit solution
\be
q - \frac{1}{3}q^3 = \frac{\alpha C^4}{(2E)^{\frac{3}{2}}} \, t \,.
\ee
The solution runs between the stationary points of the cubic
$q - \frac{1}{3}q^3$, i.e. between $q = \pm 1$, which is where
$\alpha x = \lambda$. Over a finite time interval, $\alpha x$ decreases from
$\lambda$, climbs the potential and stops at $t=0$, then increases
back to $\lambda$. The stopping point is where $q=0$, i.e. where
$\alpha x = \lambda - \frac{E^2}{C^4}$. The motion may or
may not pass through $x=0$, depending on the energy.

In practice, this model is only valid for $x$ near zero,
because it ignores the third mode entering the continuum. Also,
the squared frequency is only approximately linear in $x$. The
calculated dependence of both squared frequencies on $x$ is shown
in Fig. 1 of ref.\cite{AGMM}. This shows the crossover of the mode
frequencies at $x=0$.

In summary, to model a 2-vortex collision with the second shape
mode excited, in the adiabatic approximation, one should ignore the
third shape mode entirely until the separation reduces to
$r=\lambda/\alpha$, then for smaller $r$ use the model above, where
the two upper modes are coupled. If the
collision is considerably away from head-on, then the third mode will be only
slightly excited, the vortices will scatter, and when $r$ becomes larger than
$\lambda/\alpha$, the third mode can again be ignored. Excitation of
the second mode generates a repulsive potential, which increases the
scattering angle relative to that for purely geodesic motion. On the other
hand, in a head-on collision, the second mode converts entirely into the
third mode if the vortices reach coincidence at $r=0$. If they do, and scatter
through $90^\circ$, then the oscillations of the third mode can hit the
spectral wall where that mode enters the continuum, and it is not
clear what happens next; the simplified models we have proposed will
break down, and a full field-theory simulation of the 2-vortex collision
is probably necessary. Finally, in a collision that is near to, but
not exactly head-on, it is necessary to solve the equations for the
coupled second and third modes numerically, as the coupling becomes
strong when the frequencies become close to degenerate. We have not
investigated this.

\subsection{Numerical results}

As we observed in the previous subsection, during a head-on collision
of two vortices the second and third modes interchange, but only the
second mode (the out-of-phase superposition
of the radial mode on each vortex) can be excited if the
vortices are initially well separated. Our model for vortex
dynamics with the second and third mode excited
can be extended to the regime of well separated vortices. There it
has a form very similar to the lowest-mode case,
\be
L = \half \Omega(x,y) ({\dot x}^2 + {\dot y}^2) + \half {\dot\eta}^2 - \half
\omega_2^2(x,y) \eta^2 \,,
\label{Lag-second}
\ee
where $\eta$ is now the amplitude of the second mode and the squared
frequency $\omega_2^2$ is a monotonically decreasing function of the vortex
separation parameter $\rho$, approximately given by the expression
(\ref{secondfreq}).

\begin{figure}
\center
\includegraphics[width=0.4\columnwidth]{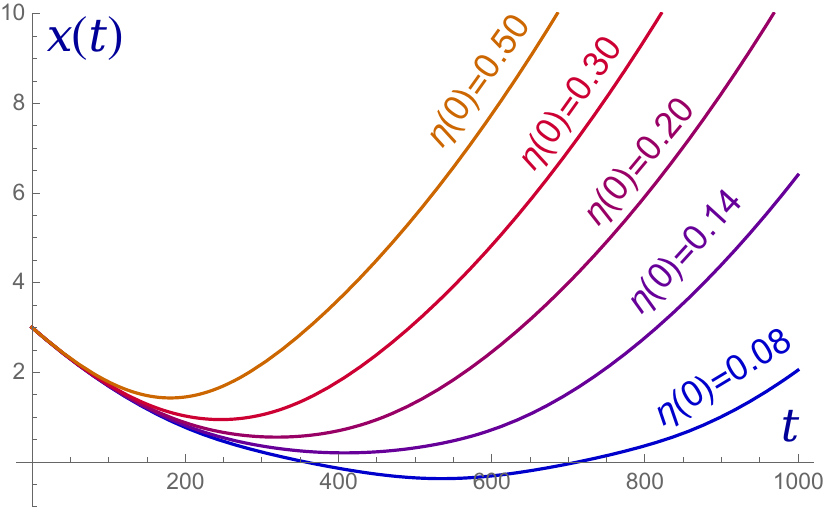}
\hspace{0.3cm}\includegraphics[width=0.4\columnwidth]{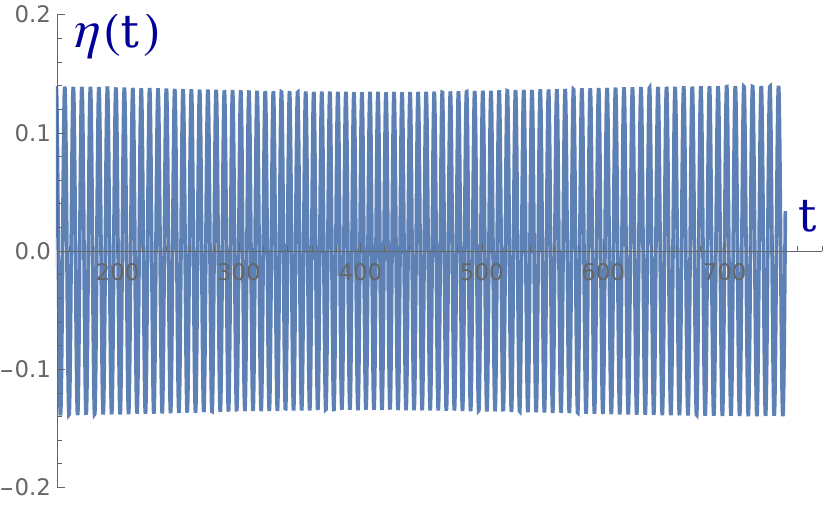}
\caption{Left: trajectories of the vortex position $x(t)$ for the
model with the second mode excited, for various values of the
initial mode amplitude, and $v_{\rm in}=0.015$. Right: evolution of
the mode amplitude $\eta(t)$ for the motion with the initial
amplitude $\eta(0)=0.14$. }
\label{repulsiv-plot}
\end{figure}
For head-on collisions we can consistently put $y=0$.
Then the Lagrangian (\ref{Lag-second}) reduces to
\be
L = \half \Omega(x) {\dot x}^2+ \half {\dot\eta}^2 - \half
\omega_2^2(x) \eta^2 \,,
\label{Lag-second-2}
\ee
with 
\be
\omega_2^2(x)=
\omega^2_2(0) - (\omega^2_2(0)-\omega^2_2(\infty))
\left(1-e^{-0.2 \sqrt{108} \, x} \right). 
\ee
Because $\omega_2^2$ increases as $x$ decreases from a positive value
and becomes negative, the inter-vortex force is always towards positive
$x$, which has the following dynamical consequences for vortices approaching
from $x=3$. If the initial amplitude of the second mode is sufficiently
large, the vortices scatter back before coalescing, i.e. they do not reach
$x=0$. With a smaller amplitude, they may pass through $x=0$ and
reach a negative $x$ before scattering back, i.e. the vortices scatter
from the horizontal to the vertical axis, stop and return to the horizontal
axis. This is a 2-bounce solution. These possibilities are plotted in
Fig. \ref{repulsiv-plot} where $v_{\rm in}=0.015$. If the amplitude is
smaller still, the vortices separate sufficiently along the vertical
axis that the third mode enters the continuum spectrum, and the
so-called spectral wall phenomenon can be expected to occur \cite{AORW}.
This possibility is not taken into account in our collective
coordinate model. 

To conclude, from this adiabatic modelling we expect to observe at
most 2-bounce scatterings of vortices in the full field theory dynamics
when the second mode is initially excited, and we do not expect any
chaotic multi-bounce pattern.

\section{Outlook}

In the present paper we proposed collective coordinate models for
the dynamics of BPS 2-vortex solutions excited by their
shape modes. The models generalize the standard geodesic flow on
the 2-vortex moduli space $\M$ by including the shape mode amplitudes
as additional collective coordinates. Importantly, in contrast to the
force-free geodesic dynamics of unexcited BPS vortices, excitation
of the shape modes introduces inter-vortex forces whose sign depends
on how the relevant mode's frequency varies over $\M$. This
can significantly change the dynamics, leading to a complete
breakdown of the original geodesic flow.

The most striking result is the appearance of a chaotic, probably
self-similar, fractal-like pattern of multi-bounce windows in
vortex-vortex collisions if the lowest shape mode is excited. This is
explained in terms of the well-known resonant energy transfer
mechanism. During collisions, the kinetic energy of vortex motion
may be temporary transferred into shape mode energy. The vortices may
then be unable to overcome the attractive interaction triggered by
the mode excitation, and instead collide once again. This process
repeats until the kinetic energy is again sufficiently large for
the vortices to separate. A similar mechanism is very well
understood in kink-antikink collisions in (1+1) dimensions, but here,
for the first time, we have shown that it can also influence the
dynamics of higher-dimensional solitons.

In particular, we have found that an even number of collisions
(bounces) changes the famous $90^\circ$ scattering of vortices
into $180^\circ$ backward scattering. Backward scattering can also occur if the
higher-frequency (second) mode is excited, because in this case
the excitation triggers a repulsive force between the vortices.
Additionally, due to level crossing, this force becomes attractive
after the vortices pass through the circularly symmetric
configuration in a head-on collision (one bounce), which makes a
second bounce more likely, resulting in backward scattering. 
No further bounces are possible in this channel.

Of course, all our predictions need to be verified by simulations of 
excited 2-vortex scattering in the full field theory \cite{Rees}. We
expect to find qualitative rather than quantitative agreement with
the collective coordinate models, because the
models make several simplifications which for a chaotic system may
not be negligible. For example, we used simplified, analytical expressions
for both the metric function and the dependence of the frequencies
on the vortex separation. Also, we did not take into account a possible
modification of the moduli space metric due to the amplitudes of the
modes, even though it is known that vibrational (massive) degrees of
freedom can affect the metric of kinetic, zero modes, see e.g. the
description of a single, vibrating kink \cite{MORW}. Such a metric
modification is obviously a subleading effect in comparison with
the appearance of an inter-vortex force, but nonetheless, it can
affect the locations of bounce windows. Also, inclusion of non-quadratic
terms in the effective potential may slightly change the dynamics.

Looking from a wider perspective, it is fascinating that effects
previously associated only with one-dimensional kinks, find their
counterparts in the dynamics of higher-dimensional solitons.

\section*{Appendix: Field theory derivation of the
effective models}

In this appendix, we sketch the derivation of the two effective models,
identified in the main sections of this paper, governing the dynamics
of BPS 2-vortices when shapes modes are excited. The
Abelian Higgs field theory action is \cite{ManSut}
\be
S=\int \left\{ -\frac{1}{4} F_{\mu\nu} F^{\mu\nu} + \frac{1}{2}
\overline{D_\mu \phi} \, D^\mu \phi - \frac{1}{8}(1-\bar\phi\phi)^2
\right\} \, dx_1dx_2 \, dt
= \int \left\{ T - V \right\} \, dt \,,
\ee
such that the kinetic and potential energy of the system in the
temporal gauge $A_0 = 0$ are, respectively,
\begin{equation}
T =\half \int \left\{ \partial_t A_i \partial_t A_i +
\partial_t \bar\phi \, \partial_t \phi \right\} \, dx_1dx_2 \,, \quad 
V = \half\int \left\{ F_{ij}F_{ij} +
\overline{D_i \phi} D_i \phi +  \frac{1}{8}(1-\bar\phi\phi)^2
\right\} \, dx_1dx_2 \,.
\end{equation}
It is well known that BPS 2-vortex solutions satisfy the
first-order equations 
\begin{equation}
F_{12}=\frac{1}{2}(1-\bar{\phi}\phi) \,, \quad D_1\phi+i D_2\phi=0 \,,
\end{equation}
and describe configurations where two unit magnetic flux vortices
have an arbitrary separation. In the center of mass
frame the 2-vortex solution is denoted as \cite{AGMM}
\begin{equation} \label{configu}
\widetilde{A}_j(X_1,X_2,x_1,x_2) \,, \quad
\widetilde{\phi}(X_1,X_2,x_1,x_2) \,,
\end{equation}
where $(X_1,X_2)$ are the spatial coordinates of a general point in
the space-time. $\pm (x_1,x_2)$ specify the
locations of the constituent vortices (zeros of the scalar field
$\widetilde{\phi}$) and play the role of real coordinates on the centred
2-vortex moduli space $\M$. They can be combined into the complex coordinate
$w=x_1+ix_2=\rho e^{i\theta}$ introduced in section 2, although as we
have seen it is more convenient to work with a complex coordinate $z=x+iy = r
e^{i\varphi}$ proportional to $w^2$, that is, $z=k w^2$. Then
$r=k\rho^2$ and $\varphi=2\theta$. Note that the angle $\theta$ varies
only in the range $[-\half\pi, \half\pi]$ because adding $\pi$ moves
the unit vortex at $(x_1,x_2)$ to $(-x_1,-x_2)$ and vice versa, but
these two configurations are identical. On the other hand,
$\varphi = 2 \theta$ has the normal angular range. In terms of the
real moduli space coordinates $(x,y)$, the 2-vortex configurations
(\ref{configu}) can be assembled as the column vector
\begin{equation}
\widetilde{\Phi}(X_1,X_2,x,y)=\left(\widetilde{A}_1(X_1,X_2,x,y) ,
\widetilde{A}_2(X_1,X_2,x,y), \widetilde{\phi}_1(X_1,X_2,x,y),
\widetilde{\phi}_2(X_1,X_2,x,y)\right)^{\rm T} \,,
\end{equation}
where $\widetilde{\phi}_1$ and $\widetilde{\phi}_2$ are the real and
imaginary parts of $\widetilde{\phi}$. 

\subsection*{Collective coordinate model for 2-vortex
dynamics with the lowest-frequency shape mode excited}

The configuration for moving vortices
with the lowest-frequency shape mode excited is written as
\begin{equation}
\widetilde{\Psi}(X_1,X_2, x(t), y(t))=\widetilde{\Phi}(X_1,X_2,x(t),y(t))
+\eta(t) \, \xi_1(X_1,X_2,x(t),y(t)) \,,
\label{adshmod}
\end{equation}
where $\xi_1$ denotes the lowest (normalized) shape mode of the
second-order fluctuation operator ${\cal H}$,
\begin{equation}
{\cal H} \, \xi_1(X_1,X_2,x,y)=\omega_1^2(r) \, \xi_1(X_1,X_2,x,y) \,, 
\label{spect1}
\end{equation} 
and $\eta(t)$ is the real shape mode amplitude. Note that the
frequency $\omega_1$ depends on the point in the moduli space where
the spectral problem is studied, but because of rotational symmetry
it only depends on $r = \sqrt{x^2 + y^2}$. The flow over the moduli space of
both the eigenfunction $\xi_1$ and the eigenvalue $\omega_1^2$ has been
numerically described in \cite{AGMM}. $x$, $y$ and $\eta$ are the
collective coordinates, whose effective dynamics is
implemented by assuming that the temporal dependence of the fields
in $\mathbb{R}^2$ is only through $\dot x$, $\dot y$ and $\dot\eta$. Under
this hypothesis, the effective dynamical system is constructed as
follows. Neglecting contributions of the form
$\eta \int \dot{\widetilde{\Phi}}^{\rm T}\dot{\xi}_1 \, dX_1dX_2 $,
$\eta \int \dot{\xi}_1^{\rm T} \dot{\widetilde{\Phi}} \, dX_1dX_2 $
and $\eta^2 \int \dot{\xi}_1^{\rm T} \dot{\xi}_1 \, dX_1dX_2 $, the
effective kinetic energy reads
\begin{equation}
T_{\rm eff}=
\frac{1}{2}g_{xx}(x,y)\dot{x}^2+\frac{1}{2}g_{yy}(x,y)\dot{y}^2
+\frac{1}{2}\dot{\eta}^2 \,,
\end{equation}
where the metric factors are
\begin{eqnarray}
&& g_{xx}(x,y)=\int
\Big(\frac{\partial {\widetilde\Phi}}{\partial x}\Big)^{\rm T} \Big(
\frac{\partial {\widetilde\Phi}}{\partial x}\Big) \, dX_1dX_2 =
\Omega(r) \nonumber \\
&& g_{yy}(x,y)=\int
\Big(\frac{\partial {\widetilde\Phi}}{\partial y}\Big)^{\rm T} \Big(
\frac{\partial {\widetilde\Phi}}{\partial y}\Big) \, dX_1dX_2 =
\Omega(r) \,.
\end{eqnarray}
These integrals define the conformal factor
$\Omega(r)$ of the Samols metric \cite{Samols}. We finally obtain
\begin{equation}
T_{\rm eff}= \frac{1}{2} \Omega(r) \Big(\dot{x}^2+\dot{y}^2
\Big)+\frac{1}{2}\dot{\eta}^2 \,.
\end{equation}
The contribution to the Lagrangian which does not depend on
time derivatives is evaluated up to second-order in $\eta$, and is
\begin{equation}
\frac{1}{2} 
\eta^2 \, \int
\xi_1^{\rm T}(X_1,X_2,x,y) \, {\cal H} \, \xi_1(X_1,X_2,x,y) \, dX_1dX_2 \,.
\end{equation}
Using (\ref{spect1}) and taking into account the rotational symmetry,
we find the effective potential over the moduli space,
\begin{equation}
V_{\rm eff} (x,y,\eta) = \frac{1}{2} \omega_1^2(r) \eta^2 \,.
\end{equation}
Because $\omega_1^2(r)$ has its minimum value $\omega_1^2(0)$ at the origin 
and increases to a finite value $\omega_1^2(\infty)
> \omega_1^2(0)$ at infinity, attractive forces arise between
the vortices when the lowest shape mode is excited. Moreover,
this potential is proportional to the squared amplitude
of the mode, opening the door to transfer of energy from the mode into
the kinetic energy of vortices moving through the moduli space.
Thus, the effective Lagrangian
\begin{equation}
L_{\rm eff} = T_{\rm eff}-V_{\rm eff}
= \frac{1}{2} \Omega(r) \Big(\dot{x}^2+\dot{y}^2
\Big)+\frac{1}{2}\dot{\eta}^2 - \frac{1}{2} \omega_1^2(r) \eta^2 \,,
\end{equation}
the starting point of section 3, is obtained from the field theory as an
effective collective coordinate model. We stress finally that
incorporating the effect of the shape mode up to second-order
generalizes to the quantized theory as a 1-loop/semi-classical
correction to moduli space dynamics.

\subsection*{Collective coordinate model for 2-vortex dynamics
with the two higher-frequency shape modes excited}

Consider the configuration
\begin{equation}
\Sigma(X_1,X_2,x(t),y(t))=\widetilde{\Phi}(X_1,X_2,x(t),y(t)) + \eta_2(t)
\, \xi_2(X_1,X_2,x(t),y(t))+\eta_3(t) \, \xi_3(X_1,X_2,x(t),y(t)) \,,
\end{equation}
which describes a dynamical 2-vortex excited by the shape modes $\xi_2$ and
$\xi_3$ having the higher frequencies $\omega_2$ and $\omega_3 \ge
\omega_2$, respectively, and amplitudes $\eta_2$
and $\eta_3$. These modes are mutually orthogonal eigenfunctions
of the second-order fluctuation operator ${\cal H}$, see \cite{AGMM},
\begin{equation}
{\cal H} \, \xi_i(X_1,X_2,x,y)=\omega_i^2(r) \,\xi_i(X_1,X_2,x,y) \,,
\hspace{0.3cm} i=2,3 \,.
\end{equation}
The spectral flow of $\omega_2^2$ and $\omega_3^2$
over $\M$ is summarized as follows:

\begin{itemize}

\item The frequencies are rotationally symmetric over $\M$ and only
depend on $r$.
  
\item At the apex of $\M$, the two frequencies degenerate:
$\omega_2^2(0)=\omega_3^2(0)$.
  
\item $\omega_2^2(r)$ decreases linearly with $r$ near $r=0$,
and approaches the limit $\omega_2^2(\infty) = \omega_1^2(\infty) > 0$.
  
\item $\omega_3^2(r)$ increases linearly with $r$ near $r=0$,
reaching 1 at the boundary of a disc in $\M$ where the mode
$\xi_3$ disappears into the continuum.
  
\end{itemize}
Arguing exactly as in the previous derivation of the effective model
for the lower-frequency shape mode we envisage the following effective
Lagrangian for the upper-frequency modes:
\begin{equation}
L_{\rm eff}
=\frac{1}{2}\Omega(r)(\dot{x}^2+\dot{y}^2)
+\frac{1}{2}\dot{\eta}_2^2+\frac{1}{2}\dot{\eta}_3^2
-\frac{1}{2}\omega_2^2(r)\eta_2^2 -
\frac{1}{2}\omega_3^2(r)\eta_3^2 \,.
\end{equation}

Investigation of the effective dynamics including the higher-frequency
modes is particularly worthwhile inside the disc where both
modes are present. Here, one can approximate any BPS
2-vortex solution by adding a zero mode of suitable amplitude to the
circularly-symmetric, coincident 2-vortex solution. This produces a
splitting of the double zero of the Higgs field into two single zeros
with a small separation $2\rho$. The collective coordinate
procedure prescribes that the motion through $\M$ is only due to
the time-dependence of $\rho(t)$. If the shape modes of frequencies
$\omega_2$ and $\omega_3$ are also excited, we are led to study the
effective dynamics of the configurations
\begin{eqnarray}
\Sigma(X_1,X_2,x(t),y(t)) &=& \widetilde{\Phi}(X_1,X_2,0,0) + \epsilon(t) \,
\xi_0(X_1,X_2,0,0)) \nonumber \\
&& \qquad + \eta_2(t) \, \xi_2(X_1,X_2,0,0)
+ \eta_3(t) \, \xi_3(X_1,X_2,0,0) \,.
\label{expansion}
\end{eqnarray}
$\epsilon$, $\eta_2$ and $\eta_3$ are, respectively, the amplitudes of
the zero mode and these two shape modes, the new collective coordinates.
Plugging the expression (\ref{expansion}) into the second-order action
we obtain
\begin{equation}
S^{(2)} = \half \int \Big\{ \|\xi_0\|^2 {\dot\epsilon}^2(t)
+ \|\xi_2\|^2 {\dot \eta}^2_2(t)
+  \|\xi_3\|^2 {\dot\eta}^2_3(t)  - \omega_2^2(r)  \|\xi_2\|^2 \eta_2^2(t)
- \omega_3^2 (r) \|\xi_3\|^2 \eta_3^2(t) \Big\} \, dt \,, 
\end{equation}
where the orthogonality of the eigenfunctions $\xi_i$ of ${\cal H}$ has
been employed. We can assume that the shape modes $\xi_2$ and $\xi_3$
are normalized, but we shall use the non-normalized zero mode
\be \label{zeromode}
\xi_0(X_1,X_2,0,0) = R \left( h(R) \sin \Theta \,, h(R) \cos
\Theta \,, -\frac{h'(R)}{f_2(R)} \,, \, 0 \right)^{\rm T}
\ee
described in \cite{AlGarGuil2,AGMM}, where $(R,\Theta)$ here denote
spatial polar coordinates, i.e. $X_1=R\cos \Theta$ and
$X_2=R\sin \Theta$. This choice of the zero mode involves the
splitting of the vortex zeros in the $x_2$-direction (vertical direction)
away from the apex of the moduli space, such that $x_2(t)$ determines
the inter-vortex distance. The expression (\ref{zeromode}) allows us
to find a relation between the amplitude $\epsilon$ of the zero mode
$\xi_0$ and $x_2$ as\footnote{The relevant modes are those with
label $(2,0)$ in ref.\cite{AlGarGuil2}.}
\be
\epsilon(t) = \frac{(\delta_2)^2}{2|c_2|} \, (x_2(t))^2 = C_1 \,
(x_2(t))^2 = \frac{C_1}{k} \, r(t)
\label{epsilon-d}
\ee
where $\delta_2\approx 0.236146$ and $c_2 \approx -0.277308$. These
values come from the local behaviour of the coincident 2-vortex Higgs
field profile $f_2(R)\approx \delta_2 R^2$ and of the zero mode
(\ref{zeromode}) with $h(R) \approx 1+c_2 R^2$. Additionally,
\begin{equation}
\omega_2^2 (r) = \lambda - \frac{C_2}{k} \, r \,, \hspace{0.5cm}
\omega_3^2 (r) = \lambda + \frac{C_2}{k} \, r
\label{omega-d}
\end{equation}
where $\lambda=\omega_2^2(0) = \omega_3^2(0) \approx 0.97303$ and
$C_2 \approx 0.025873$. Working with the expressions (\ref{epsilon-d})
and (\ref{omega-d}), the action becomes
\be
S^{(2)} = \half \int \Bigg\{  \frac{C_1^2}{k^2} \|\xi_0\|^2 \,
\dot{r}^2(t)  + \dot{\eta}_2^2(t) + \dot{\eta}_3^2(t)  
- \left( \lambda -\frac{C_2}{k} \, r(t) \right) \,
\eta_2^2(t) - \left( \lambda + \frac{C_2}{k} r(t) \right) \,
\eta_3^2(t) \Bigg\} \, dt \,,
\label{s2}   
\ee
with $\|\xi_0\|^2 \approx 62.4936$. We now combine the two real shape
mode amplitudes into a single complex amplitude $\chi$,
\begin{equation}
\eta_3=\frac{1}{\sqrt{2}}
\left(e^{\frac{i}{2}\varphi}\chi^*+e^{-\frac{i}{2}\varphi}\chi \right)
=\eta^*_3 \,, \quad \eta_2=\frac{1}{\sqrt{2} \, i}
\left(e^{\frac{i}{2}\varphi}\chi^* -e^{-\frac{i}{2}\varphi}\chi \right)
=\eta^*_2 \,,
\end{equation}
which involves the argument $\varphi$ of the coordinate
$z=r e^{i\varphi}$ on the moduli space. This induces a rotation which
generalizes the dynamics. Now the motion can be in any direction
through the origin of the moduli space, and the Lagrangian in
(\ref{s2}) becomes
\be
L_{\rm eff} = \maketitle\frac{1}{2} \frac{C_1^2}{k^2} \|\xi_0\|^2 \,
\dot{z}^*\dot{z} + \dot{\chi}^*\dot{\chi} - \lambda \chi^*\chi
- \half \frac{C_2}{k} z^* \chi^2 - \half \frac{C_2}{k} z \chi^{*2} \,.
\ee
Finally, if the value of $k$ is set as $k=C_1 \|\xi_0\|$ and we define
$\alpha= C_2/(C_1 \|\xi_0\|)$ then the effective Lagrangian becomes
\be
L_{\rm eff} = \maketitle\frac{1}{2} \, \dot{z}^*\dot{z}
+ \dot{\chi}^*\dot{\chi} - \lambda \chi^*\chi - \half \alpha z^* \chi^2
- \half \alpha z \chi^{*2} \,,
\ee
which reproduces the expression (\ref{realLag}) introduced in the main text.

\section*{Acknowledgments}

This research was supported by the Spanish MCIN with funding from European
Union NextGenerationEU (PRTRC17.I1) and Consejeria de Educacion from
JCyL through the QCAYLE project, as well as the MCIN project
PID2020-113406GB-I0. This research has made use of the
high-performance computing resources of the Castilla y Le\'on
Supercomputing Center (SCAYLE), financed by the European Regional
Development Fund (ERDF). NSM is partially supported by UK STFC consolidated
grant ST/T000694/1. AW was supported by the Polish National Science
Centre, grant NCN 2019/35/B/ST2/00059.

We acknowledge Morgan Rees for informing us of results concerning
the scattering of excited vortices prior to publication.

\end{document}